\documentclass[12pt]{article}
\usepackage{sectsty}
\usepackage{geometry}
\usepackage{graphicx}
\usepackage{indentfirst} 
\usepackage[english]{babel}
\usepackage[utf8x]{inputenc}
\usepackage{amsmath}
\usepackage{autobreak}
\usepackage{booktabs}
\usepackage{longtable}
\usepackage{array}
\usepackage{multirow}
\usepackage{wrapfig}
\usepackage{float}
\usepackage{colortbl}
\usepackage{pdflscape}
\usepackage{tabu}
\usepackage{threeparttable}
\usepackage{threeparttablex}
\usepackage[normalem]{ulem}
\usepackage{makecell}
\usepackage{graphicx}
\usepackage{setspace}
\usepackage{cite}
\usepackage{authblk}

\title{Testing Homogeneity of Proportion Ratios for Stratified Bilateral Correlated Data} 

\author[a]{Wanqing Tian}
\author[b]{Chang-Xing Ma \thanks{CONTACT Chang-Xing Ma: cxma@buffalo.edu, Department of Biostatistics, University at Buffalo, 3435 Main St., Buffalo, NY 14214, USA}}
\affil[a,b]{Department of Biostatistics, University at Buffalo, Buffalo, New York, USA}

\sectionfont{\small}
\subsectionfont{\small}
\subsubsectionfont{\small}

\begin{document}
\maketitle

\textbf{Abstract} 

Intraclass correlation in bilateral data has been investigated in recent decades with various statistical methods. In practice, stratifying bilateral data by some control variables will provide more sophisticated statistical results to satisfy different research proposed in random clinical trials. In this article, we propose three test statistics (likelihood ratio test, score test, and Wald-type test statistics) to evaluate the homogeneity of proportion ratios for stratified bilateral correlated data under an equal correlation assumption. Monte Carlo simulations of Type I error and power are performed, and the score test yields a robust outcome based on empirical Type I error and power. Lastly, a real data example is conducted to illustrate the proposed three tests.


\section{Introduction}


In random clinical trials, bilateral data frequently occurs in patients who receive the treatment based on paired body parts or organs (such as eyes, ears, kidneys, and so on). Since the outcome of bilateral data has been naturally split into three kinds: no response, unilateral response, and bilateral responses, considering the intraclass correlation in the bilateral data is a natural way to avoid misleading results \cite{Rosner_1982, Donner1989rhoModel, Dallal_1988, ma2017rho, Ma_2013R}. Intraclass correlation in bilateral data has been investigated in the recent decades with various statistical methods \cite{Rosner_1982, Donner1989rhoModel, Dallal_1988, Ma_2013R, Tang_2013matchedpair, Tang_Tang_Qiu_2008, Liu_2016, Shen_2019OR}. Rosner proposes that a conditional probability of response at one side of paired body parts or organs gives a response at the other body parts or organs is a positive constant R time the response rate  \cite{Rosner_1982}. Tang et al. provide a statistical inference for correlated data in binary paired data under R model, and also evaluate the asymptotic test with Type I error and power \cite{Tang_2006}. Ma, Shan, and Liu develop an asymptotic testing method under homogeneity assumption for Rosner’s R model \cite{Ma_2013R}. Donner assumes that all treatment groups share one intra-class correlation coefficient $\rho$  \cite{Donner1989rhoModel}, and Thompson evaluates the robustness of Donnar’s $\rho$ model in pair data by adopting simulations \cite{Thompson_1993}. Liu et al. test equality of correlation coefficients based on Donner’s $\rho$ model for paired binary data with multiple groups  \cite{liu2016testing}. Later, Liu et al. explore the exact methods of testing the homogeneity of prevalence for correlated binary data under Donnar’s $\rho$ model  \cite{Liu_2016}. However, Dallal criticizes Rosner’s assumption and points out that “the constant R model will give a poor fit if the characteristic is almost certain to occur bilaterally with widely varying group-specific prevalence” \cite{Dallal_1988}. Dallal believes the conditional probability of response at one side of paired body parts or organs, giving a response at the other body parts or organs is a constant $\gamma$ \cite{Dallal_1988}. Li et al. develop asymptotic and exact methods following the Dallal’s model  \cite{li2020statistical}. Then, Chen et al. propose multiple test statistics of response rates in the different groups under the Dallal’s model  \cite{chen2022further}.

The homogeneity test for the appropriate effect size measure between different groups equals to test the common value of the measure of effect. Moreover, there are three popular methods to evaluate the effect size in random clinical trials: odds ratio, risk difference, and relative risk   \cite{newcombe2012confidence}. Indeed, relative risk is more informative than risk different in some cases \cite{wang2015exact}. Compared with odds ratio, the relative risk can process sparse data better. The stratifying of bilateral data by some control variables (e.g.  disease phases, age, etc.) will provide more sophisticated statistical results to satisfy different research proposed in random clinical trials \cite{Peng_2019, Tang_2011, Tang_2012}. For example, evaluating the appropriate effect size across strata of disease phases between treatment and control groups equals to test the effect size of homogeneity across the strata. Zhuang et al. investigate the homogeneity test of ratio of two proportions in the stratified bilateral data based on the Donner’s $\rho$ model \cite{zhuang2018homogeneity}. Shen et al. conduct the testing homogeneity of difference of two proportions for stratified correlated paired binary data under Donner’s $\rho$ model \cite{shen2017testing}. Xue and Ma propose interval estimation of proportion ratios for stratified bilateral correlated binary data under Rosner’s constant R model \cite{Xue_2019SMMR}.

The primary goal of this article is to investigate the proportion ratios to clinical trial design with stratified bilateral data under Dallal’s model. The remainder of this paper is organized as follows: section 2 presents the data structure and hypotheses, and section 3 introduces the maximum-likelihood estimation under homogeneity. We propose three tests to examine the homogeneity of proportions across strata under Dallal’s model in section 4. Accordingly, we investigate the performance and robustness of three tests by using simulation studies in section 5. In section 6, we use a real example of the OME study to illustrate our proposed methods. Conclusions and future works are in section 7.

\section{Data structure and Hypotheses}

\subsection{Notation}

Let $m_{l i j}$ be the number of patients in the $i^{t h}$ group of the $j^{t h}$ stratum with $l^{t h}$ responses, where $i$ = 1,2, $j = 1, \ldots,J$ , $l$ = 0, 1, 2. $m_{+i j}=m_{0 i j}+m_{1 i j}+m_{2 i j}$ represent the number of patients in the $i^{t h}$ group of the $j^{t h}$ stratum. The $p_{l i j}$ is corresponding to the probability of patients in  $i^{t h}$ group of the $j^{t h}$ stratum with $l^{t h}$ responses. Define $Z_{i j k h}$ as the indicator of the response of $k^{t h}$ eyes of the $h^{t h}$ patient in the $i^{t h}$ group from $j^{t h}$ stratum, where $k$ = 1, 2 and $h=1,2, \ldots, m_{+i j}$. If $Z_{i j k h} = 1$ then the improvement response occurs, and  $Z_{i j k h} = 0$ otherwise. Therefore, donate $\operatorname{Pr}\left(Z_{h i j k}=1\right)= \pi_{i j},\left(0 \leq \pi_{i j} \leq 1\right)$ as probability of having response on one site.

$$
  \begin{array}{cccc}
\hline & \multicolumn{2}{c}{\text { Group }} \\
\cline { 2 - 3 } \text { Number of Responses }(l) & 1 & 2 & \quad \text { Total } \\
\hline 0 & m_{01 j}\left(p_{01 j}\right) & m_{02 j}\left(p_{02 j}\right) & m_{0+j} \\
1 & m_{11 j}\left(p_{11 j}\right) & m_{12 j}\left(p_{12 j}\right) & m_{1+j} \\
2 & m_{21 j}\left(p_{21 j}\right) & m_{22 j}\left(p_{22 j}\right) & m_{2+j} \\
\hline \text { Total } & m_{+1 j} & m_{+2 j} & m_{++j} \\
\hline
\end{array}
$$
  
We explore the intraclass correlation based on Dallal's model that is conditional probability of response at one side of paired body parts or organs, giving a response at the other body parts or organs is a constant $\gamma$. Therefore, we assume $P\left(Z_{h i j k}=1 \mid Z_{h i j(3-k)}=1\right)=\gamma_{j}$, and the probabilities of eye with none, one or both can be express as following:

$$p_{0 i j }=1-\left(2-\gamma_{  j}\right) \pi_{i j },$$

$$\quad p_{1 i j }=2 \pi_{i j }\left(1-\gamma_{ j}\right),$$

$$\quad p_{2 i j }=\pi_{i j} \gamma_{ j},$$
and $p_{0 i j}+p_{1 i j}+p_{2 i j}=1$ for any fixed $i$ and $j$. The joint likelihood function for the observed data $\boldsymbol{m_{i j}}=\left(m_{0 i j}, m_{1 i j}, m_{2 i j}\right)^{T}$ is given by 

$$f\left(\boldsymbol{m_{i j}} \mid \boldsymbol{p_{i j}}\right)=\prod_{j=1}^{J} \prod_{i=1}^{2} \frac{m_{+i j} !}{m_{0 i j} ! m_{1 i j} ! m_{2 i j} !} p_{0 i j}^{m_{0 i j}} p_{1 i j}^{m_{1 i j}} p_{2 i j}^{m_{2 i j}} , $$ where $\boldsymbol{p_{i j}}=\left(p_{0 i j}, p_{1 i j}, p_{2 i j}\right)^{T}$. The corresponding log-likelihood function can be expressed as:
$$
l\left( \boldsymbol{\pi_{i}}, \boldsymbol{\gamma}\right)=\sum_{j=1}^{J} \sum_{i=1}^{2}\left\{m_{0 i j} \log \left[1-\left(2-\gamma_{j}\right) \pi_{i j}\right]+m_{1 i j} \log \left[2 \pi_{i j}\left(1-\gamma_{j}\right)\right]+m_{2 i j} \log \left[\pi_{i j} \gamma_{j}\right]\right\}+\log C
, $$ where  $\boldsymbol{\pi_{i}} = \left(\pi_{i 1}, \ldots, \pi_{i J}\right)^T$, $\boldsymbol{\gamma} = \left(\gamma_{1}, \ldots, \gamma_{ J}\right)^T$, and $C=\prod_{j=1}^{J} \prod_{i=1}^{2} \frac{m_{+i j} !}{m_{0 i j} ! m_{1 i j} ! m_{2 i j} !}$ is a constant.

\subsection{Hypotheses for the proportion ratio across strata}

Assuming there is a common ratio of proportions between two groups across $J$ strata, i.e., the ratio of proportions between two groups in the $j^{\text {th }}$ stratum is $\delta_{j}=\pi_{2 j} / \pi_{1 j}$ for $j=1, \ldots, J$, the hypotheses are given as

$$
H_{0}: \delta_{1}=\delta_{2}= ...=\delta_{J} \quad \text { versus } \quad H_{a}: \delta_{1} \neq \delta_{2} \neq ... \neq \delta_{J}.
$$

\section{Maximum-likelihood estimation (MLE) under homogeneity}
\subsection{The unconstrained MLEs}

Since we assume $\delta_{j}=\pi_{2 j} / \pi_{1 j}$ for $j=1, \ldots, J$, then $\pi_{2 j} = \delta_{j} \pi_{1 j}$. Under the null hypothesis with common $\delta$, the log-likelihood ratio can be express as:

$$
l\left(\boldsymbol{\pi_1}, \delta, \boldsymbol{\gamma}\right)=\sum_{j=1}^J l_j\left(\pi_{1j}, \delta, \gamma_{j}\right),
$$
where 
$$
\begin{aligned}
\label{unconstrained_mle}
\quad l_{j}\left(\pi_{1 j}, \delta, \gamma_{j} \right)&= \{m_{01 j} \log \left[1-\left(2-\gamma_{j}\right) \pi_{1 j}\right] +m_{11 j} \log \left[2 \pi_{1 j}\left(1-\gamma_{j}\right)\right] +m_{21 j} \log \left[\pi_{1 j} \gamma_{j}\right]\\
&+m_{02 j} \log \left[1-\left(2-\gamma_{j}\right) \pi_{1 j} \delta\right]+m_{12 j} \log \left[2 \pi_{1 j} \delta\left(1-\gamma_{j}\right)\right] +m_{22 j} \log \left[\pi_{1 j} \delta \gamma_{j}\right] \}\\
&+\log C.
\end{aligned}
$$
Differentiating $ l\left(\boldsymbol{\pi_1}, \delta, \boldsymbol{\gamma}\right)$ with respect to $\boldsymbol{\pi_1}$ and $\boldsymbol{\gamma}$ , we have 

$$\frac{\partial^{} l_{}}{\partial \pi_{1 j}^{}} = \frac{ m_{1+j}}{ \pi_{1 j }} + \frac{ m_{2+j}}{ \pi_{1 j }} + \frac{ m_{01j}\,\left( \gamma_{ j}-2\right)}{ \pi_{1 j }\,\left( \gamma_{ j}-2\right)+1}+\frac{\delta \, m_{02j}\,\left( \gamma_{ j}-2\right)}{\delta \, \pi_{1 j }\,\left( \gamma_{ j}-2\right)+1},$$ 

$$\frac{\partial^{} l_{}}{\partial \gamma_{ j}^{}} = \frac{ m_{21j}}{ \gamma_{ j}}+\frac{ m_{22j}}{ \gamma_{ j}}+\frac{ m_{11j}}{ \gamma_{ j}-1}+\frac{ m_{12j}}{ \gamma_{ j}-1}+\frac{ m_{01j}\, \pi_{1 j }}{ \pi_{1 j }\,\left( \gamma_{ j}-2\right)+1}+\frac{\delta \, m_{02j}\, \pi_{1 j }}{\delta \, \pi_{1 j }\,\left( \gamma_{ j}-2\right)+1}.$$
We can set $\frac{\partial^{} l_{}}{\partial \pi_{1 j}^{}} =0 $ with $\frac{\partial^{} l_{}}{\partial \gamma_{ j}^{}} =0$ to get MLEs $\hat{\pi }_{1 j}$ and $\hat{\gamma}_j$. Indeed, MLE of $\gamma_{j}$ has closed-form solution. Meanwhile, MLE of $\pi_{1 j}$ is a function of $m$ and $\delta$. 
For the MLE of $\delta$, we will update by using Fisher scoring iterative algorithm, and 

$$\hat{\delta} = \delta^{(t+1)}=\delta^{(t)}+\left.I_{ }^{-1}\left(\delta^{(t)}\right)\left( \frac{\partial l_{ }}{\partial \delta}\right)\right|_{\delta=\delta(t), \hat{\pi }_{1 j}, \hat{\gamma_{j}} }, $$
where

$$\frac{\partial^{ } l_{}}{ \partial \delta } = \sum_{j=1}^{J} (\frac{ m_{12j} }{\delta }+\frac{ m_{22j} }{\delta }+\frac{ m_{02j} \, \pi_{1 j } \,{\left(\gamma_j -2\right)}}{\delta \, \pi_{1 j } \,{\left(\gamma_j -2\right)}+1}) , $$

$$
\frac{\partial^{2} l_{}}{ \partial \delta^{2} } =\sum_{j=1}^{J} (-\frac{ m_{12j} }{\delta^2 }-\frac{ m_{22j} }{\delta^2 }-\frac{ m_{02j} \,{ \pi_{1 j } }^2 \,{{\left(\gamma_j -2\right)}}^2 }{{{\left(\delta \, \pi_{1 j } \,{\left(\gamma_j -2\right)}+1\right)}}^2 }) ,
$$
and 
$$
\begin{aligned}
\label{I}
I = - E(\frac{\partial^{2} l_{}}{ \partial \delta^{2} }) &= -E (\sum_{j=1}^{J} (-\frac{ m_{12j} }{\delta^2 }-\frac{ m_{22j} }{\delta^2 }-\frac{ m_{02j} \,{ \pi_{1 j } }^2 \,{{\left(\gamma_j -2\right)}}^2 }{{{\left(\delta \, \pi_{1 j } \,{\left(\gamma_j -2\right)}+1\right)}}^2 })) \\
& =
\sum_{j=1}^{J} (\frac{ m_{+2j} p_{12j} }{\delta^2 }+\frac{ m_{+2j} p_{22j} }{\delta^2 }+\frac{ m_{+2j} p_{02j} \,{ \pi_{1 j } }^2 \,{{\left(\gamma_j -2\right)}}^2 }{{{\left(\delta \, \pi_{1 j } \,{\left(\gamma_j -2\right)}+1\right)}}^2 }) .
\end{aligned} 
$$

\subsection{The Global MLEs}

Under the global hypothesis, $H_{a}: \delta_{1} \neq \delta_{2} \neq ... \neq \delta_{J}$ , the log-likelihood can be presenting as:

$$
l\left(\boldsymbol{\pi_1}, \boldsymbol{\delta}, \boldsymbol{\gamma}\right)=\sum_{j=1}^J l_j\left(\pi_{1j}, \delta_{j}, \gamma_{j}\right),
$$
where
$$
\begin{aligned}
\label{globel_mle}
\quad l_{j}\left(\pi_{1 j}, \delta_{j}, \gamma_{j}\right) &=\sum_{j=1}^{J}\left\{m_{01 j} \log \left[1-\left(2-\gamma_{j}\right) \pi_{1 j}\right]\right. +m_{11 j} \log \left[2 \pi_{1 j}\left(1-\gamma_{j}\right)\right] +m_{21 j} \log \left[\pi_{1 j} \gamma_{j}\right] \\
&+m_{02 j} \log \left[1-\left(2-\gamma_{j}\right) \pi_{1 j} \delta_{j}\right]
+m_{12 j} \log \left[2 \pi_{1 j} \delta_{j}\left(1-\gamma_{j}\right)\right]
\left.+m_{22 j} \log \left[\pi_{1 j} \delta_{j} \gamma_{j}\right]\right\}+\log C ,
\end{aligned}
$$
and the MLEs of three parameters $\pi_{1 j}$, $\delta_{j}$, $\gamma_{j}$ can be derivied by setting the partial differentiation equal to zero, then the MLEs as following:

$$
\tilde {\pi}_{1 j} = \frac{{\left( m_{11j} + m_{21j} \right)}\,{\left( m_{11j} + m_{12j} +2\, m_{21j} +2\, m_{22j} \right)}}{2\,{\left( m_{01j} + m_{11j} + m_{21j} \right)}\,{\left( m_{11j} + m_{12j} + m_{21j} + m_{22j} \right)}} ,
$$

$$\tilde{\gamma}_{j} =\frac{2\, m_{2+j} }{ m_{1+j}+2\, m_{2+j} } , $$

$$\tilde{\delta_{j}} =  -\frac{{\left( m_{12j} + m_{22j} \right)}\,{\left( m_{01j} - m_{12j} + m_{1+j}- m_{22j} + m_{2+j} \right)}}{{\left( m_{02j} + m_{12j} + m_{22j} \right)}\,{\left( m_{12j} - m_{1+j}+ m_{22j} - m_{2+j} \right)}} . $$

\section{Testing methods}
\subsection{Likelihood Ratio Test ($T_{L}$) }

The likelihood ratio test statistic is given by

$$
\begin{aligned}
T_{L} &=2 \sum_{j=1}^{J}\left[l_{j}\left(\tilde{\delta_{j}}, \tilde{\pi}_{1 j}, \tilde{\gamma}_{j}\right)-l_{j}\left(\hat{\delta}, \hat{\pi }_{1 j}, \hat{\gamma}_j\right) \right], 
\end{aligned}
$$
Moreover,  $T_{L}$ is asymptotically distributed as a Chi-square distribution with $ J - 1$ degree of freedom under the null hypothesis.



\subsection{Score Test  ($T_{S C}$) }

Under the assumption that each stratum has the same ratio of proportions $\delta_{j}$ ($\delta = \delta_{j}$), the score test statistic is given by

$$
\begin{aligned}
T_{S C} &=\sum_{j=1}^{J} U_{j} I_{j}^{-1} U_{j}^{T} \mid \delta_{ }=\hat{\delta_{ }}, \pi_{1 j}=\hat{\pi }_{1 j}, \gamma_{j}=\hat{\gamma}_j \\
&=\sum_{j=1}^{J}\left(\frac{\partial l_{j}}{\partial \delta_{j}}\right)^{2} \frac{1}{D_{j}} I_{j}^{-1}(1,1) \mid \delta_{ }=\hat{\delta_{ }}, \pi_{1 j}=\hat{\pi }_{1 j}, \gamma_{j}=\hat{\gamma}_j,
\end{aligned}
$$

$$I_{j}^{-1} = \left (\frac{1}{D_{j}} \right) \left(\begin{array}{ccc}
{\left(I_{22 }^{(j)} \,I_{33 }^{(j)} -I_{23 }^{(j)} \,I_{32 }^{(j)} \right) }\, & -{\left(I_{12 }^{(j)} \,I_{33 }^{(j)} -I_{13 }^{(j)} \,I_{32 }^{(j)} \right) }\, & {\left(I_{12 }^{(j)} \,I_{23 }^{(j)} -I_{13 }^{(j)} \,I_{22 }^{(j)} \right) }\,  \\
-{\left(I_{21 }^{(j)} \,I_{33 }^{(j)} -I_{23 }^{(j)} \,I_{31 }^{(j)} \right) }\,  & {\left(I_{11 }^{(j)} \,I_{33 }^{(j)} -I_{13 }^{(j)} \,I_{31 }^{(j)} \right) }\,   & -{\left(I_{11 }^{(j)} \,I_{23 }^{(j)} -I_{13 }^{(j)} \,I_{21 }^{(j)} \right) }\, \\
{\left(I_{21 }^{(j)} \,I_{32 }^{(j)} -I_{22 }^{(j)} \,I_{31 }^{(j)} \right) }\, & -{\left(I_{11 }^{(j)} \,I_{32 }^{(j)} -I_{12 }^{(j)} \,I_{31 }^{(j)} \right) }\,   & {\left(I_{11 }^{(j)} \,I_{22 }^{(j)} -I_{12 }^{(j)} \,I_{21 }^{(j)} \right) }\,     
\end{array}\right) , $$

$$D_{j}= I_{11 }^{(j)} \,I_{22 }^{(j)} \,I_{33 }^{(j)} -I_{11 }^{(j)} \,I_{23 }^{(j)} \,I_{32 }^{(j)} -I_{12 }^{(j)} \,I_{21 }^{(j)} \,I_{33 }^{(j)} +I_{12 }^{(j)} \,I_{23 }^{(j)} \,I_{31 }^{(j)} +I_{13 }^{(j)} \,I_{21 }^{(j)} \,I_{32 }^{(j)} -I_{13 }^{(j)} \,I_{22 }^{(j)} \,I_{31 }^{(j)} , $$ where the score function for $j$ th stratum $U_{j}=\left(\frac{\partial l_{j}}{\partial \delta_{ }}, 0,0\right)$, and $I_{j}^{-1}(1,1)=\left(I_{22}^{(j)} I_{33}^{(j)}-I_{23}^{(j)} I_{32}^{(j)}\right)$ is the first diagonal element of the inverse of fisher information matrix (See Appendix for more detail). 



\subsection{ Wald-type Test ($T_{W}$) }
An alternative way to express the null hypothesis is in matrix form as $\boldsymbol{C} \boldsymbol{\beta}^{T}=0$, where $$\boldsymbol{\beta}=\left(\delta_{1}, \pi_{11}, \gamma_{1}, \delta_{2}, \pi_{12}, \gamma_{2}, \cdots, \delta_{J}, \pi_{1 J}, \gamma_{J}\right)$$ and

$$
\boldsymbol{C}=\left(\begin{array}{ccccccccccccccc}
1 & 0 & 0 & -1 & 0 & 0 & 0 & 0 & 0 & \cdots & \cdots & \cdots & 0 & 0 & 0 \\
1 & 0 & 0 & 0 & 0 & 0 & -1 & 0 & 0 & \cdots & \cdots & \cdots & 0 & 0 & 0 \\
\vdots & \vdots & \vdots & \vdots & \vdots & \vdots & 0 & 0 & 0 & \ddots & \ddots & \ddots & \vdots & \vdots & \vdots \\
1 & 0 & 0 & 0 & 0 & 0 & 0 & 0 & 0 & \cdots & \cdots & \cdots & -1 & 0 & 0
\end{array}\right) .
$$
The global information matrix for this Wald-type test $I_{W}$ is a $3 J \times 3 J$ block diagonal matrix, with each block being a $3 \times 3$ information matrix of parameter vector $\left(\delta_{j}, \pi_{1 j}, \gamma_{j}\right)$ within each stratum. We used the same derivation approach as we derive information matrix for the score test in previous.

$$
I_{W}=\left[\begin{array}{cccc}
I_{W}^{(1)} & 0 & \cdots & 0 \\
0 & I_{W}^{(2)} & \cdots & 0 \\
\vdots & \vdots & \ddots & \vdots \\
0 & 0 & \cdots & I_{W}^{(J)}
\end{array}\right]_{3 J \times 3 J  } ,  
$$
with
$$
I_{W}^{(j)}=\left[\begin{array}{ccc}
I_{W 11}^{(j)} & I_{W 12}^{(j)} & I_{W 13}^{(j)} \\
I_{W 21}^{(j)} & I_{W 22}^{(j)} & I_{W 23}^{(j)} \\
I_{W 31}^{(j)} & I_{W 32}^{(j)} & I_{W 33}^{(j)}
\end{array}\right]_{3 \times 3} ,
$$
the explicit form of each element $\left(I_{W i k}^{(j)}\right)$ being exactly the same as $\left(I_{i k}^{j}\right)$ derived in previous for unconstrained score test. Then the Wald-type test statistic can be express as: 
$$
T_{W}=\left.\left(\boldsymbol{\beta} \boldsymbol{C}^{T}\right)\left(\boldsymbol{C} I_{W}^{-1} \boldsymbol{C}^{T}\right)^{-1}\left(\boldsymbol{C} \boldsymbol{\beta}^{T}\right)\right|_{\boldsymbol{\beta}=\left(\tilde{\delta}_{1}, \ldots, \tilde{\delta}_{J}, \tilde{\pi}_{11}, \ldots, \tilde{\pi}_{1 J}, \tilde{\gamma}_{1}, \ldots, \tilde{\gamma}_{J}\right)}
$$
$$
I_{W}^{-1}=\left[\begin{array}{cccc}
\left(I_{W}^{(1)}\right)^{-1} & 0 & \cdots & 0 \\
0 & \left(I_{W}^{(2)}\right)^{-1} & \cdots & 0 \\
\vdots & \vdots & \ddots & \vdots \\
0 & 0 & \cdots & \left(I_{W}^{(J)}\right)^{-1}
\end{array}\right]_{J \times J} , 
$$
where


$$(I_{W}^{(j)})^{-1} = \left (\frac{1} {D_{W} ^{(j)}} \right) \left(\begin{array}{ccc}
{\left(I_{W22 }^{(j)} \,I_{W33 }^{(j)} -I_{W23 }^{(j)} \,I_{W32 }^{(j)} \right) }\, & -{\left(I_{W12 }^{(j)} \,I_{W33 }^{(j)} -I_{W13 }^{(j)} \,I_{W32 }^{(j)} \right) }\, & {\left(I_{W12 }^{(j)} \,I_{W23 }^{(j)} -I_{W13 }^{(j)} \,I_{W22 }^{(j)} \right) }\,  \\
-{\left(I_{W21 }^{(j)} \,I_{W33 }^{(j)} -I_{W23 }^{(j)} \,I_{W31 }^{(j)} \right) }\,  & {\left(I_{W11 }^{(j)} \,I_{W33 }^{(j)} -I_{W13 }^{(j)} \,I_{W31 }^{(j)} \right) }\,   & -{\left(I_{W11 }^{(j)} \,I_{W23 }^{(j)} -I_{W13 }^{(j)} \,I_{W21 }^{(j)} \right) }\, \\
{\left(I_{W21 }^{(j)} \,I_{W32 }^{(j)} -I_{W22 }^{(j)} \,I_{W31 }^{(j)} \right) }\, & -{\left(I_{W11 }^{(j)} \,I_{W32 }^{(j)} -I_{W12 }^{(j)} \,I_{W31 }^{(j)} \right) }\,   & {\left(I_{W11 }^{(j)} \,I_{W22 }^{(j)} -I_{W12 }^{(j)} \,I_{W21 }^{(j)} \right) }\,      
\end{array}\right) , $$

$$D_{W} ^{(j)}= I_{W11 }^{(j)} \,I_{W22 }^{(j)} \,I_{W33 }^{(j)} -I_{W11 }^{(j)} \,I_{W23 }^{(j)} \,I_{W32 }^{(j)} -I_{W12 }^{(j)} \,I_{W21 }^{(j)} \,I_{W33 }^{(j)} +I_{W12 }^{(j)} \,I_{W23 }^{(j)} \,I_{W31 }^{(j)} +I_{W13 }^{(j)} \,I_{W21 }^{(j)} \,I_{W32 }^{(j)} -I_{W13 }^{(j)} \,I_{W22 }^{(j)} \,I_{W31 }^{(j)} . $$

\section{Simulation Studies}

This section investigates the empirical performances of three proposed test statistics in the previous section by three Monte Carlo simulation studies to evaluate the quality of relative risk in terms of the empirical Type I error rate and the power.

\subsection{Empirical Type I error rates}

In the first Monte Carlo simulation study, we investigate the behavior of empirical Type I rate for three proposed tests under various procedures, where $m$ = 25, 50, or 100 in strata $J$ = 2, 4, 6, or 8. By considering $\pi_{1 j}$ and $\gamma_{j}$ are either common or different cross strata,
we provide the parameter settings under different sample sizes and various sets of parameters in Table 1. For each configuration, 50,000 samples are randomly generated under null hypothesis $H_{0}: \delta = 1, 1.2, 0.8$, and the empirical Type I error rates are calculated by dividing the number of rejections by 50,000. According to the previous study \cite{Xue_2019SMMR}, the robustness of empirical Type 1 error rates is in the range of 0.04 to 0.06. As shown in Table 2-5, we observe that the Wald-type test has poor performance, and the likelihood ratio test does not work well under a small sample with multiple strata. However, the score test is more robust than the likelihood ratio and Wald-type test.


\begin{singlespace}
\begin{longtable}[t]{llllll}
\caption{\label{tab:unnamed-chunk-6}Parameter setups for computing empirical Type I error rates and powers.}\\
\toprule
\multicolumn{1}{c}{} & \multicolumn{1}{c}{} & \multicolumn{4}{c}{Number of strata} \\
\cmidrule(l{3pt}r{3pt}){3-6}
Parameter & Cases & J=2 & J=4 & J=6 & J=8\\
\midrule
\endfirsthead
\caption[]{Parameter setups for computing empirical Type I error rates and powers. \textit{(continued)}}\\
\toprule
\multicolumn{1}{c}{} & \multicolumn{1}{c}{} & \multicolumn{4}{c}{Number of strata} \\
\cmidrule(l{3pt}r{3pt}){3-6}
Parameter & Cases & J=2 & J=4 & J=6 & J=8\\
\midrule
\endhead

\endfoot
\bottomrule
\endlastfoot
 & I & (0.2,0.4) & (0.2,0.4,0.2,0.4) & (0.2,0.4,0.2,0.4,0.2,0.4) & (0.2,0.4,0.2,0.4,0.2,0.4,0.2,0.4)\\
$\gamma$ & II & (0.3,0.3) & (0.3,0.3,0.3,0.3) & (0.3,0.3,0.3,0.3,0.3,0.3) & (0.3,0.3,0.3,0.3,0.3,0.3,0.3,0.3)\\
 & III & (0.3,0.5) & (0.3,0.5,0.3,0.5) & (0.3,0.5,0.3,0.5,0.3,0.5) & (0.3,0.5,0.3,0.5,0.3,0.5,0.3,0.5)\\
 & IV & (0.6,0.6) & (0.6,0.6,0.6,0.6) & (0.6,0.6,0.6,0.6,0.6,0.6) & (0.6,0.6,0.6,0.6,0.6,0.6,0.6,0.6)\\
 &  &  &  &  & \\

$\pi_{1}$ & a & (0.2,0.4) & (0.2,0.4,0.2,0.4) & (0.2,0.4,0.2,0.4,0.2,0.4) & (0.2,0.4,0.2,0.4,0.2,0.4,0.2,0.4)\\
 & b & (0.3,0.3) & (0.3,0.3,0.3,0.3) & (0.3,0.3,0.3,0.3,0.3,0.3) & (0.3,0.3,0.3,0.3,0.3,0.3,0.3,0.3)\\
 & c & (0.2,0.3) & (0.2,0.3,0.2,0.3) & (0.2,0.3,0.2,0.3,0.2,0.3) & (0.2,0.3,0.2,0.3,0.2,0.3,0.2,0.3)\\*
\end{longtable}
\end{singlespace}

\begin{singlespace}
\begin{longtable}[t]{rllrrrrrrrrr}
\caption{\label{tab:unnamed-chunk-7}Simulation results of the empirical sizes for 2 strata.}\\
\toprule
\multicolumn{1}{c}{} & \multicolumn{1}{c}{} & \multicolumn{1}{c}{} & \multicolumn{3}{c}{m = 25} & \multicolumn{3}{c}{m = 50} & \multicolumn{3}{c}{m = 100} \\
\cmidrule(l{3pt}r{3pt}){4-6} \cmidrule(l{3pt}r{3pt}){7-9} \cmidrule(l{3pt}r{3pt}){10-12}
$\delta$ & $\gamma$ & $\pi_{1}$ & $T_{L}$ & $T_{SC}$ & $T_{W}$ & $T_{L}$ & $T_{SC}$ & $T_{W}$ & $T_{L}$ & $T_{SC}$ & $T_{W}$\\
\midrule
\endfirsthead
\caption[]{Simulation results of the empirical sizes for 2 strata. \textit{(continued)}}\\
\toprule
\multicolumn{1}{c}{} & \multicolumn{1}{c}{} & \multicolumn{1}{c}{} & \multicolumn{3}{c}{m = 25} & \multicolumn{3}{c}{m = 50} & \multicolumn{3}{c}{m = 100} \\
\cmidrule(l{3pt}r{3pt}){4-6} \cmidrule(l{3pt}r{3pt}){7-9} \cmidrule(l{3pt}r{3pt}){10-12}
$\delta$ & $\gamma$ & $\pi_{1}$ & $T_{L}$ & $T_{SC}$ & $T_{W}$ & $T_{L}$ & $T_{SC}$ & $T_{W}$ & $T_{L}$ & $T_{SC}$ & $T_{W}$\\
\midrule
\endhead

\endfoot
\bottomrule
\endlastfoot
1.0 & I & a & 5.46 & 5.28 & 3.76 & 5.28 & 5.17 & 3.97 & 5.06 & 5.03 & 4.41\\
 &  & b & 5.39 & 5.37 & 2.04 & 5.28 & 5.23 & 3.40 & 5.09 & 5.07 & 4.20\\
 &  & c & 5.31 & 5.25 & 1.71 & 5.19 & 5.15 & 2.99 & 5.13 & 5.11 & 4.02\\
 & II & a & 5.65 & 5.43 & 4.68 & 5.35 & 5.23 & 4.45 & 5.27 & 5.22 & 4.73\\
 &  & b & 5.42 & 5.42 & 1.92 & 5.39 & 5.33 & 3.33 & 4.94 & 4.92 & 4.05\\

 &  & c & 5.54 & 5.46 & 2.39 & 5.22 & 5.16 & 3.23 & 5.10 & 5.08 & 4.00\\
 & III & a & 5.36 & 5.21 & 3.42 & 5.33 & 5.24 & 3.92 & 5.14 & 5.10 & 4.27\\
 &  & b & 5.50 & 5.47 & 1.83 & 5.21 & 5.15 & 3.14 & 5.07 & 5.04 & 4.12\\
 &  & c & 5.41 & 5.29 & 1.52 & 5.14 & 5.11 & 2.79 & 5.11 & 5.09 & 3.87\\
 & IV & a & 5.51 & 5.21 & 3.95 & 5.12 & 5.01 & 3.77 & 5.03 & 4.97 & 4.23\\

 &  & b & 5.49 & 5.44 & 1.20 & 5.22 & 5.20 & 2.79 & 4.94 & 4.90 & 3.78\\
 &  & c & 5.48 & 5.34 & 1.70 & 5.32 & 5.20 & 2.71 & 5.24 & 5.18 & 3.73\\
1.2 & I & a & 5.37 & 5.22 & 3.66 & 5.24 & 5.15 & 4.00 & 4.98 & 4.95 & 4.29\\
 &  & b & 5.54 & 5.48 & 1.70 & 5.16 & 5.10 & 3.15 & 5.06 & 5.04 & 4.08\\
 &  & c & 5.56 & 5.47 & 1.48 & 5.25 & 5.18 & 2.76 & 5.12 & 5.11 & 3.87\\

 & II & a & 5.42 & 5.29 & 4.47 & 5.11 & 5.06 & 4.28 & 5.14 & 5.10 & 4.70\\
 &  & b & 5.19 & 5.15 & 1.51 & 5.05 & 5.00 & 3.03 & 5.00 & 4.98 & 4.01\\
 &  & c & 5.49 & 5.44 & 2.17 & 5.11 & 5.07 & 3.15 & 5.07 & 5.04 & 3.91\\
 & III & a & 5.25 & 5.13 & 3.32 & 5.08 & 4.99 & 3.63 & 5.09 & 5.05 & 4.27\\
 &  & b & 5.28 & 5.24 & 1.46 & 5.14 & 5.09 & 2.99 & 5.13 & 5.10 & 4.00\\

 &  & c & 5.53 & 5.46 & 1.31 & 5.29 & 5.25 & 2.56 & 4.88 & 4.86 & 3.65\\
 & IV & a & 5.37 & 5.12 & 3.76 & 5.17 & 5.05 & 3.87 & 5.20 & 5.12 & 4.29\\
 &  & b & 5.62 & 5.57 & 1.05 & 5.12 & 5.09 & 2.48 & 5.26 & 5.24 & 3.84\\
 &  & c & 5.64 & 5.50 & 1.44 & 5.14 & 5.09 & 2.46 & 5.17 & 5.14 & 3.55\\
0.8 & I & a & 5.31 & 5.13 & 3.81 & 5.16 & 5.07 & 4.14 & 5.15 & 5.09 & 4.47\\

 &  & b & 5.36 & 5.33 & 2.49 & 5.15 & 5.10 & 3.60 & 5.10 & 5.06 & 4.34\\
 &  & c & 5.54 & 5.39 & 2.08 & 5.17 & 5.12 & 3.12 & 5.08 & 5.04 & 4.10\\
 & II & a & 5.40 & 5.14 & 4.63 & 5.20 & 5.08 & 4.34 & 5.09 & 5.03 & 4.63\\
 &  & b & 5.41 & 5.36 & 2.30 & 5.30 & 5.24 & 3.68 & 5.10 & 5.08 & 4.29\\
 &  & c & 5.29 & 5.14 & 2.45 & 5.25 & 5.18 & 3.39 & 4.99 & 4.95 & 4.00\\

 & III & a & 5.45 & 5.28 & 3.77 & 5.20 & 5.11 & 3.95 & 5.12 & 5.06 & 4.42\\
 &  & b & 5.31 & 5.28 & 2.02 & 5.25 & 5.22 & 3.49 & 5.00 & 4.99 & 4.14\\
 &  & c & 5.29 & 5.18 & 1.75 & 5.26 & 5.17 & 2.95 & 5.19 & 5.12 & 3.98\\
 & IV & a & 5.72 & 5.43 & 4.07 & 5.23 & 5.12 & 3.99 & 5.12 & 5.08 & 4.39\\
 &  & b & 5.43 & 5.34 & 1.55 & 5.30 & 5.26 & 3.02 & 5.01 & 4.98 & 3.84\\

 &  & c & 5.51 & 5.19 & 1.95 & 5.23 & 5.12 & 2.82 & 5.34 & 5.29 & 3.86\\*
\end{longtable}
\end{singlespace}


\begin{singlespace}
\begin{longtable}[t]{rllrrrrrrrrr}
\caption{\label{tab:unnamed-chunk-7}Simulation results of the empirical sizes for 4 strata.}\\
\toprule
\multicolumn{1}{c}{} & \multicolumn{1}{c}{} & \multicolumn{1}{c}{} & \multicolumn{3}{c}{m = 25} & \multicolumn{3}{c}{m = 50} & \multicolumn{3}{c}{m = 100} \\
\cmidrule(l{3pt}r{3pt}){4-6} \cmidrule(l{3pt}r{3pt}){7-9} \cmidrule(l{3pt}r{3pt}){10-12}
$\delta$ & $\gamma$ & $\pi_{1}$ & $T_{L}$ & $T_{SC}$ & $T_{W}$ & $T_{L}$ & $T_{SC}$ & $T_{W}$ & $T_{L}$ & $T_{SC}$ & $T_{W}$\\
\midrule
\endfirsthead
\caption[]{Simulation results of the empirical sizes for 4 strata. \textit{(continued)}}\\
\toprule
\multicolumn{1}{c}{} & \multicolumn{1}{c}{} & \multicolumn{1}{c}{} & \multicolumn{3}{c}{m = 25} & \multicolumn{3}{c}{m = 50} & \multicolumn{3}{c}{m = 100} \\
\cmidrule(l{3pt}r{3pt}){4-6} \cmidrule(l{3pt}r{3pt}){7-9} \cmidrule(l{3pt}r{3pt}){10-12}
$\delta$ & $\gamma$ & $\pi_{1}$ & $T_{L}$ & $T_{SC}$ & $T_{W}$ & $T_{L}$ & $T_{SC}$ & $T_{W}$ & $T_{L}$ & $T_{SC}$ & $T_{W}$\\
\midrule
\endhead

\endfoot
\bottomrule
\endlastfoot
1.0 & I & a & 5.82 & 5.45 & 4.33 & 5.33 & 5.17 & 4.35 & 5.07 & 5.03 & 4.59\\
 &  & b & 5.41 & 5.20 & 2.44 & 5.21 & 5.09 & 3.29 & 5.13 & 5.08 & 4.04\\
 &  & c & 5.73 & 5.46 & 2.49 & 5.31 & 5.20 & 3.19 & 5.11 & 5.05 & 4.00\\
 & II & a & 5.52 & 5.13 & 4.96 & 5.23 & 5.10 & 4.78 & 5.12 & 5.03 & 4.92\\
 &  & b & 5.54 & 5.28 & 2.41 & 5.23 & 5.12 & 3.27 & 5.12 & 5.07 & 4.06\\

 &  & c & 5.63 & 5.31 & 2.89 & 5.29 & 5.15 & 3.30 & 5.13 & 5.04 & 4.09\\
 & III & a & 5.75 & 5.41 & 3.79 & 5.26 & 5.08 & 4.11 & 5.06 & 5.01 & 4.53\\
 &  & b & 5.52 & 5.30 & 2.39 & 5.05 & 4.93 & 3.14 & 5.05 & 5.01 & 3.91\\
 &  & c & 5.69 & 5.35 & 2.22 & 5.35 & 5.20 & 3.20 & 5.15 & 5.10 & 3.81\\
 & IV & a & 5.59 & 5.16 & 4.05 & 5.23 & 5.00 & 4.55 & 5.19 & 5.10 & 4.64\\

 &  & b & 5.55 & 5.30 & 1.84 & 5.23 & 5.12 & 2.85 & 5.16 & 5.09 & 3.72\\
 &  & c & 5.77 & 5.33 & 2.43 & 5.36 & 5.18 & 3.01 & 5.15 & 5.08 & 3.75\\
1.2 & I & a & 5.67 & 5.34 & 3.67 & 5.26 & 5.10 & 4.16 & 5.17 & 5.08 & 4.51\\
 &  & b & 5.65 & 5.40 & 2.05 & 5.23 & 5.13 & 3.18 & 5.12 & 5.07 & 3.90\\
 &  & c & 5.65 & 5.35 & 1.96 & 5.26 & 5.13 & 2.90 & 5.21 & 5.14 & 3.74\\

 & II & a & 5.77 & 5.41 & 4.60 & 5.19 & 5.09 & 4.55 & 4.97 & 4.89 & 4.70\\
 &  & b & 5.54 & 5.32 & 1.92 & 5.34 & 5.19 & 2.96 & 5.20 & 5.17 & 4.02\\
 &  & c & 5.71 & 5.37 & 2.37 & 5.32 & 5.18 & 3.33 & 5.04 & 4.95 & 3.96\\
 & III & a & 5.56 & 5.24 & 3.56 & 5.21 & 5.03 & 3.85 & 5.27 & 5.17 & 4.41\\
 &  & b & 5.45 & 5.21 & 1.86 & 5.41 & 5.26 & 3.07 & 5.17 & 5.11 & 3.85\\

 &  & c & 5.53 & 5.22 & 1.82 & 5.32 & 5.20 & 2.74 & 5.38 & 5.32 & 3.81\\
 & IV & a & 5.72 & 5.26 & 4.00 & 5.30 & 5.13 & 4.11 & 4.96 & 4.88 & 4.38\\
 &  & b & 5.47 & 5.19 & 1.56 & 5.15 & 5.04 & 2.48 & 5.23 & 5.18 & 3.61\\
 &  & c & 5.78 & 5.37 & 1.92 & 5.29 & 5.13 & 2.84 & 5.41 & 5.30 & 3.65\\
0.8 & I & a & 5.56 & 5.23 & 4.86 & 5.24 & 5.03 & 4.66 & 5.16 & 5.07 & 4.71\\

 &  & b & 5.50 & 5.21 & 2.98 & 5.03 & 4.92 & 3.59 & 5.09 & 5.01 & 4.25\\
 &  & c & 5.77 & 5.34 & 2.98 & 5.24 & 5.06 & 3.37 & 5.24 & 5.16 & 4.18\\
 & II & a & 5.60 & 5.14 & 5.68 & 5.21 & 5.04 & 5.18 & 4.94 & 4.88 & 4.84\\
 &  & b & 5.53 & 5.23 & 2.87 & 5.37 & 5.23 & 3.64 & 5.14 & 5.08 & 4.22\\
 &  & c & 5.73 & 5.33 & 3.47 & 5.20 & 5.01 & 3.73 & 5.24 & 5.17 & 4.37\\

 & III & a & 5.64 & 5.19 & 4.57 & 5.42 & 5.25 & 4.60 & 5.04 & 4.95 & 4.62\\
 &  & b & 5.63 & 5.34 & 2.96 & 5.43 & 5.27 & 3.76 & 5.09 & 5.04 & 4.19\\
 &  & c & 5.80 & 5.36 & 2.82 & 5.39 & 5.20 & 3.46 & 5.19 & 5.10 & 4.04\\
 & IV & a & 5.83 & 5.24 & 5.10 & 5.38 & 5.15 & 4.90 & 5.15 & 5.06 & 4.78\\
 &  & b & 5.66 & 5.30 & 2.35 & 5.41 & 5.23 & 3.24 & 5.13 & 5.06 & 3.96\\

 &  & c & 5.75 & 5.12 & 3.02 & 5.33 & 5.07 & 3.36 & 5.02 & 4.89 & 3.79\\*
\end{longtable}
\end{singlespace}


\begin{singlespace}
\begin{longtable}[t]{rllrrrrrrrrr}
\caption{\label{tab:unnamed-chunk-7}Simulation results of the empirical sizes for 6 strata.}\\
\toprule
\multicolumn{1}{c}{} & \multicolumn{1}{c}{} & \multicolumn{1}{c}{} & \multicolumn{3}{c}{m = 25} & \multicolumn{3}{c}{m = 50} & \multicolumn{3}{c}{m = 100} \\
\cmidrule(l{3pt}r{3pt}){4-6} \cmidrule(l{3pt}r{3pt}){7-9} \cmidrule(l{3pt}r{3pt}){10-12}
$\delta$ & $\gamma$ & $\pi_{1}$ & $T_{L}$ & $T_{SC}$ & $T_{W}$ & $T_{L}$ & $T_{SC}$ & $T_{W}$ & $T_{L}$ & $T_{SC}$ & $T_{W}$\\
\midrule
\endfirsthead
\caption[]{Simulation results of the empirical sizes for 6 strata. \textit{(continued)}}\\
\toprule
\multicolumn{1}{c}{} & \multicolumn{1}{c}{} & \multicolumn{1}{c}{} & \multicolumn{3}{c}{m = 25} & \multicolumn{3}{c}{m = 50} & \multicolumn{3}{c}{m = 100} \\
\cmidrule(l{3pt}r{3pt}){4-6} \cmidrule(l{3pt}r{3pt}){7-9} \cmidrule(l{3pt}r{3pt}){10-12}
$\delta$ & $\gamma$ & $\pi_{1}$ & $T_{L}$ & $T_{SC}$ & $T_{W}$ & $T_{L}$ & $T_{SC}$ & $T_{W}$ & $T_{L}$ & $T_{SC}$ & $T_{W}$\\
\midrule
\endhead

\endfoot
\bottomrule
\endlastfoot
1.0 & I & a & 5.90 & 5.47 & 4.77 & 5.30 & 5.11 & 4.55 & 5.06 & 4.99 & 4.63\\
 &  & b & 5.56 & 5.25 & 2.67 & 5.32 & 5.17 & 3.39 & 5.23 & 5.15 & 4.19\\
 &  & c & 5.89 & 5.49 & 2.80 & 5.37 & 5.17 & 3.31 & 5.17 & 5.07 & 3.95\\
 & II & a & 5.72 & 5.20 & 5.61 & 5.33 & 5.10 & 5.00 & 5.03 & 4.92 & 4.85\\
 &  & b & 5.57 & 5.27 & 2.54 & 5.50 & 5.32 & 3.45 & 5.11 & 5.05 & 3.97\\

 &  & c & 5.58 & 5.15 & 3.07 & 5.33 & 5.12 & 3.63 & 5.25 & 5.16 & 4.33\\
 & III & a & 5.66 & 5.19 & 4.50 & 5.31 & 5.08 & 4.37 & 5.29 & 5.20 & 4.71\\
 &  & b & 5.79 & 5.41 & 2.57 & 5.33 & 5.17 & 3.40 & 5.14 & 5.05 & 4.00\\
 &  & c & 6.03 & 5.59 & 2.73 & 5.45 & 5.23 & 3.30 & 5.08 & 4.98 & 3.88\\
 & IV & a & 5.98 & 5.38 & 5.21 & 5.38 & 5.13 & 4.72 & 5.14 & 5.01 & 4.73\\

 &  & b & 5.97 & 5.58 & 2.38 & 5.26 & 5.07 & 3.06 & 5.18 & 5.09 & 3.87\\
 &  & c & 5.91 & 5.27 & 2.93 & 5.31 & 5.04 & 3.40 & 5.26 & 5.15 & 3.91\\
1.2 & I & a & 5.79 & 5.34 & 4.14 & 5.28 & 5.04 & 4.24 & 5.22 & 5.13 & 4.55\\
 &  & b & 5.75 & 5.39 & 2.27 & 5.40 & 5.24 & 3.23 & 5.13 & 5.07 & 4.01\\
 &  & c & 5.56 & 5.19 & 2.32 & 5.60 & 5.44 & 3.24 & 5.24 & 5.16 & 3.78\\

 & II & a & 5.93 & 5.43 & 4.87 & 5.27 & 5.00 & 4.59 & 5.12 & 4.98 & 4.75\\
 &  & b & 5.83 & 5.47 & 2.11 & 5.14 & 4.97 & 3.02 & 5.11 & 5.03 & 3.92\\
 &  & c & 5.79 & 5.37 & 2.74 & 5.54 & 5.32 & 3.55 & 5.30 & 5.19 & 4.08\\
 & III & a & 5.81 & 5.32 & 3.92 & 5.36 & 5.14 & 4.13 & 5.25 & 5.10 & 4.35\\
 &  & b & 5.68 & 5.33 & 2.15 & 5.37 & 5.23 & 3.05 & 5.15 & 5.10 & 3.81\\

 &  & c & 5.79 & 5.36 & 2.06 & 5.28 & 5.08 & 2.96 & 5.31 & 5.20 & 3.72\\
 & IV & a & 5.96 & 5.42 & 4.48 & 5.21 & 4.92 & 4.37 & 5.27 & 5.17 & 4.62\\
 &  & b & 5.77 & 5.42 & 1.77 & 5.36 & 5.16 & 2.81 & 5.27 & 5.19 & 3.66\\
 &  & c & 5.74 & 5.21 & 2.28 & 5.58 & 5.32 & 3.17 & 5.36 & 5.26 & 3.84\\
0.8 & I & a & 5.97 & 5.44 & 5.71 & 5.23 & 4.99 & 5.01 & 5.21 & 5.11 & 5.17\\

 &  & b & 5.81 & 5.41 & 3.51 & 5.41 & 5.24 & 4.04 & 5.21 & 5.11 & 4.32\\
 &  & c & 6.06 & 5.46 & 3.69 & 5.21 & 4.99 & 3.65 & 5.15 & 5.06 & 4.29\\
 & II & a & 5.88 & 5.29 & 6.69 & 5.29 & 5.07 & 5.64 & 5.00 & 4.87 & 5.15\\
 &  & b & 5.69 & 5.32 & 3.41 & 5.28 & 5.09 & 3.83 & 5.02 & 4.93 & 4.21\\
 &  & c & 5.93 & 5.38 & 4.21 & 5.36 & 5.13 & 4.17 & 5.18 & 5.09 & 4.53\\

 & III & a & 6.00 & 5.39 & 5.73 & 5.29 & 5.09 & 4.87 & 5.31 & 5.22 & 4.99\\
 &  & b & 5.90 & 5.52 & 3.49 & 5.25 & 5.07 & 3.67 & 5.32 & 5.21 & 4.41\\
 &  & c & 6.05 & 5.44 & 3.63 & 5.26 & 5.06 & 3.62 & 5.26 & 5.16 & 4.22\\
 & IV & a & 6.24 & 5.46 & 6.31 & 5.43 & 5.09 & 5.45 & 5.35 & 5.19 & 5.09\\
 &  & b & 5.93 & 5.44 & 3.03 & 5.35 & 5.15 & 3.46 & 5.10 & 4.98 & 4.10\\

 &  & c & 6.14 & 5.34 & 4.21 & 5.43 & 5.11 & 3.93 & 5.24 & 5.10 & 4.18\\*
\end{longtable}


\begin{longtable}[t]{rllrrrrrrrrr}
\caption{\label{tab:unnamed-chunk-7}Simulation results of the empirical sizes for 8 strata.}\\
\toprule
\multicolumn{1}{c}{} & \multicolumn{1}{c}{} & \multicolumn{1}{c}{} & \multicolumn{3}{c}{m = 25} & \multicolumn{3}{c}{m = 50} & \multicolumn{3}{c}{m = 100} \\
\cmidrule(l{3pt}r{3pt}){4-6} \cmidrule(l{3pt}r{3pt}){7-9} \cmidrule(l{3pt}r{3pt}){10-12}
$\delta$ & $\gamma$ & $\pi_{1}$ & $T_{L}$ & $T_{SC}$ & $T_{W}$ & $T_{L}$ & $T_{SC}$ & $T_{W}$ & $T_{L}$ & $T_{SC}$ & $T_{W}$\\
\midrule
\endfirsthead
\caption[]{Simulation results of the empirical sizes for 8 strata. \textit{(continued)}}\\
\toprule
\multicolumn{1}{c}{} & \multicolumn{1}{c}{} & \multicolumn{1}{c}{} & \multicolumn{3}{c}{m = 25} & \multicolumn{3}{c}{m = 50} & \multicolumn{3}{c}{m = 100} \\
\cmidrule(l{3pt}r{3pt}){4-6} \cmidrule(l{3pt}r{3pt}){7-9} \cmidrule(l{3pt}r{3pt}){10-12}
$\delta$ & $\gamma$ & $\pi_{1}$ & $T_{L}$ & $T_{SC}$ & $T_{W}$ & $T_{L}$ & $T_{SC}$ & $T_{W}$ & $T_{L}$ & $T_{SC}$ & $T_{W}$\\
\midrule
\endhead

\endfoot
\bottomrule
\endlastfoot
1.0 & I & a & 5.87 & 5.39 & 5.01 & 5.40 & 5.15 & 4.72 & 5.02 & 4.91 & 4.68\\
 &  & b & 5.66 & 5.28 & 2.79 & 5.37 & 5.16 & 3.58 & 5.17 & 5.08 & 4.08\\
 &  & c & 6.04 & 5.51 & 3.06 & 5.40 & 5.19 & 3.68 & 5.16 & 5.06 & 4.03\\
 & II & a & 5.98 & 5.33 & 5.96 & 5.47 & 5.16 & 5.43 & 5.11 & 4.97 & 5.02\\
 &  & b & 5.62 & 5.19 & 2.64 & 5.36 & 5.19 & 3.54 & 5.13 & 5.04 & 3.99\\

 &  & c & 5.67 & 5.08 & 3.61 & 5.40 & 5.16 & 3.78 & 5.03 & 4.94 & 4.06\\
 & III & a & 5.81 & 5.31 & 4.85 & 5.42 & 5.19 & 4.65 & 5.11 & 4.98 & 4.52\\
 &  & b & 5.80 & 5.38 & 2.83 & 5.13 & 4.94 & 3.24 & 5.24 & 5.15 & 4.06\\
 &  & c & 5.97 & 5.48 & 2.96 & 5.44 & 5.21 & 3.44 & 5.31 & 5.21 & 4.11\\
 & IV & a & 5.90 & 5.21 & 5.68 & 5.53 & 5.23 & 5.07 & 5.31 & 5.16 & 4.90\\

 &  & b & 6.07 & 5.62 & 2.38 & 5.23 & 5.02 & 3.10 & 5.19 & 5.10 & 3.90\\
 &  & c & 6.19 & 5.45 & 3.45 & 5.44 & 5.18 & 3.63 & 5.20 & 5.08 & 4.04\\
1.2 & I & a & 5.95 & 5.37 & 4.26 & 5.37 & 5.12 & 4.38 & 5.27 & 5.14 & 4.72\\
 &  & b & 5.76 & 5.27 & 2.31 & 5.50 & 5.28 & 3.27 & 5.31 & 5.21 & 3.97\\
 &  & c & 5.84 & 5.36 & 2.35 & 5.39 & 5.15 & 3.15 & 5.43 & 5.33 & 4.06\\

 & II & a & 6.09 & 5.40 & 5.15 & 5.49 & 5.20 & 5.09 & 5.19 & 5.06 & 4.95\\
 &  & b & 5.72 & 5.29 & 2.12 & 5.20 & 4.98 & 3.02 & 5.13 & 5.03 & 3.90\\
 &  & c & 6.03 & 5.49 & 3.03 & 5.43 & 5.23 & 3.40 & 5.21 & 5.11 & 4.07\\
 & III & a & 5.86 & 5.22 & 3.99 & 5.34 & 5.01 & 4.19 & 5.30 & 5.15 & 4.52\\
 &  & b & 5.71 & 5.28 & 2.18 & 5.40 & 5.26 & 3.11 & 5.18 & 5.10 & 4.00\\

 &  & c & 5.80 & 5.31 & 2.37 & 5.49 & 5.29 & 3.02 & 5.33 & 5.23 & 3.93\\
 & IV & a & 5.82 & 5.22 & 4.93 & 5.43 & 5.15 & 4.62 & 5.12 & 4.98 & 4.63\\
 &  & b & 5.59 & 5.17 & 1.80 & 5.30 & 5.11 & 2.80 & 5.38 & 5.30 & 3.75\\
 &  & c & 6.01 & 5.38 & 2.58 & 5.52 & 5.23 & 3.30 & 5.25 & 5.10 & 3.83\\
0.8 & I & a & 6.10 & 5.48 & 6.34 & 5.39 & 5.12 & 5.46 & 5.16 & 5.01 & 5.15\\

 &  & b & 5.85 & 5.43 & 3.84 & 5.40 & 5.19 & 4.12 & 5.14 & 5.02 & 4.47\\
 &  & c & 5.83 & 5.23 & 4.21 & 5.35 & 5.08 & 4.11 & 5.04 & 4.94 & 4.23\\
 & II & a & 5.92 & 5.28 & 7.28 & 5.47 & 5.24 & 5.95 & 5.30 & 5.19 & 5.45\\
 &  & b & 5.77 & 5.32 & 3.69 & 5.33 & 5.14 & 3.94 & 5.20 & 5.11 & 4.25\\
 &  & c & 6.12 & 5.41 & 4.79 & 5.42 & 5.17 & 4.40 & 5.18 & 5.07 & 4.61\\

 & III & a & 5.94 & 5.33 & 5.89 & 5.45 & 5.18 & 5.44 & 5.08 & 4.99 & 4.92\\
 &  & b & 5.79 & 5.28 & 3.77 & 5.56 & 5.32 & 4.04 & 5.23 & 5.13 & 4.43\\
 &  & c & 6.03 & 5.32 & 4.22 & 5.45 & 5.13 & 3.99 & 5.44 & 5.28 & 4.36\\
 & IV & a & 6.30 & 5.38 & 7.31 & 5.48 & 5.12 & 5.70 & 5.20 & 5.04 & 5.29\\
 &  & b & 5.77 & 5.25 & 3.46 & 5.46 & 5.23 & 3.73 & 5.33 & 5.23 & 4.25\\

 &  & c & 6.35 & 5.36 & 4.69 & 5.48 & 5.14 & 4.36 & 5.37 & 5.18 & 4.51\\*
\end{longtable}
\end{singlespace}

To obtain completed and robust empirical performances of three proposed tests, we propose the second simulation by randomly choosing parameter configurations. We randomly generate 1000 parameter configurations within parameter space for different sample sizes $m$ = 15, 25, 50, 100 and variety of strata $J$ = 2, 4, 6 and 8 for 50,000 replications. Same as the first simulation, the empirical Type I error rates can be calculated as the number of rejections divided by 50,000. Furthermore, the corresponding boxplots and violin plots are shown in Figures 1, 2, 3, and 4. Comparing empirical Type I error rates among three tests, the score and likelihood ratio tests are stable and robust under all conditions. Still, the Wald-type test performs poorly compared to the other two tests.

\begin{figure}[htp]
    \centering
    \includegraphics[width=16cm]{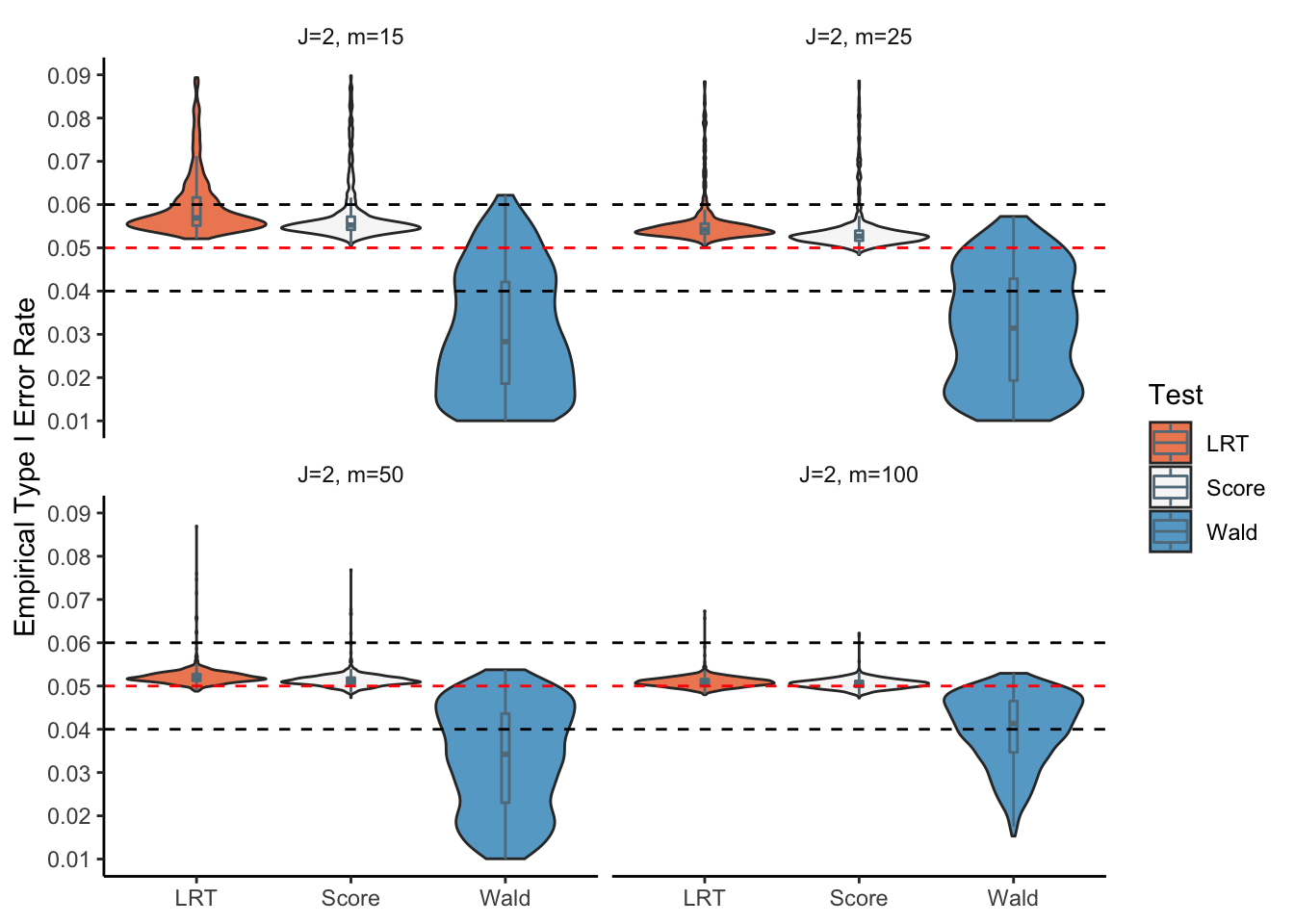}
    \caption{Violin-plots and Boxplots of empirical sizes (J=2)}
    \label{fig:galaxy}
\end{figure}

\begin{figure}[htp]
    \centering
    \includegraphics[width=16cm]{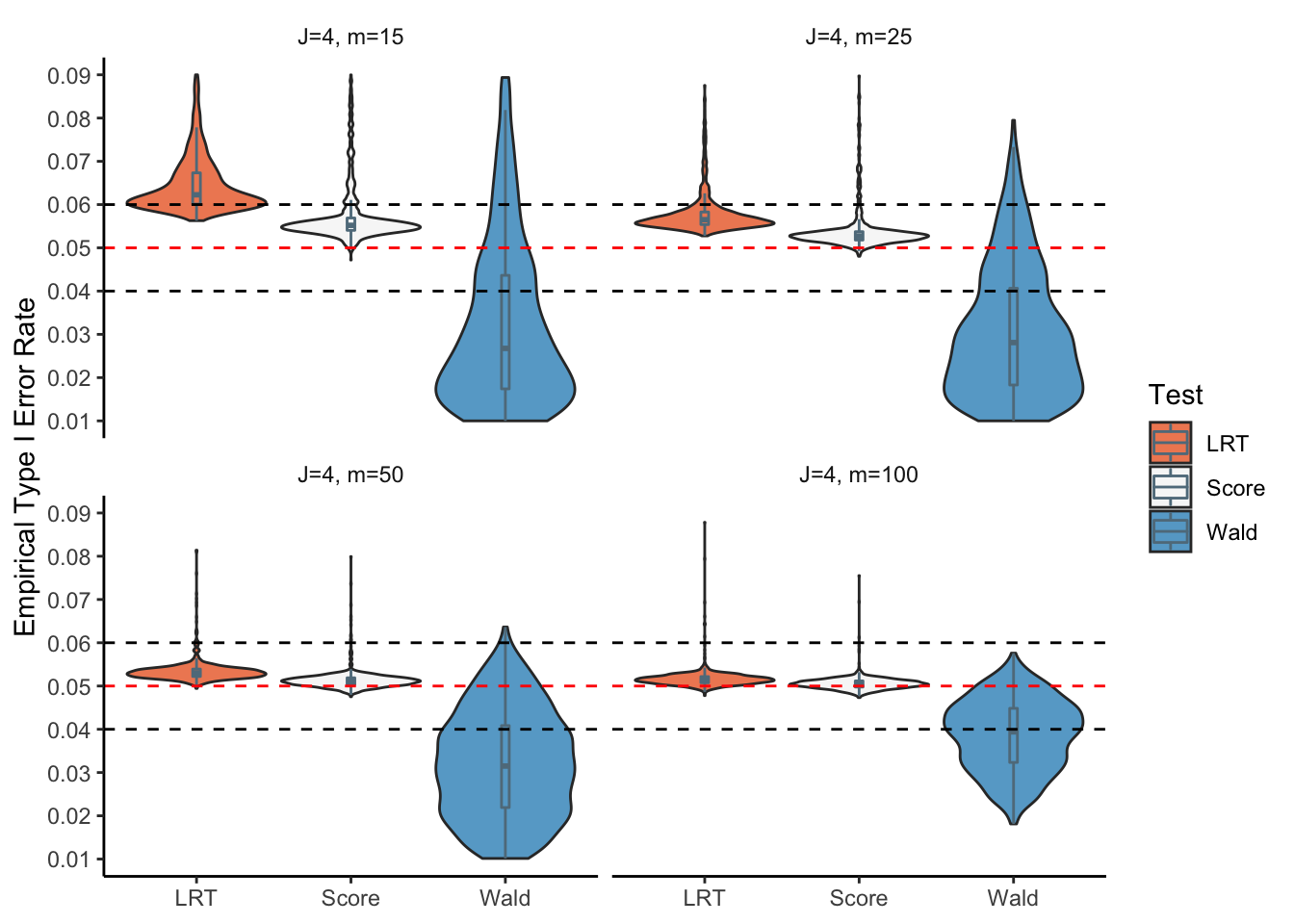}
    \caption{Violin-plots and Boxplots of empirical sizes (J=4)}
    \label{fig:galaxy}
\end{figure}

\begin{figure}[htp]
    \centering
    \includegraphics[width=16cm]{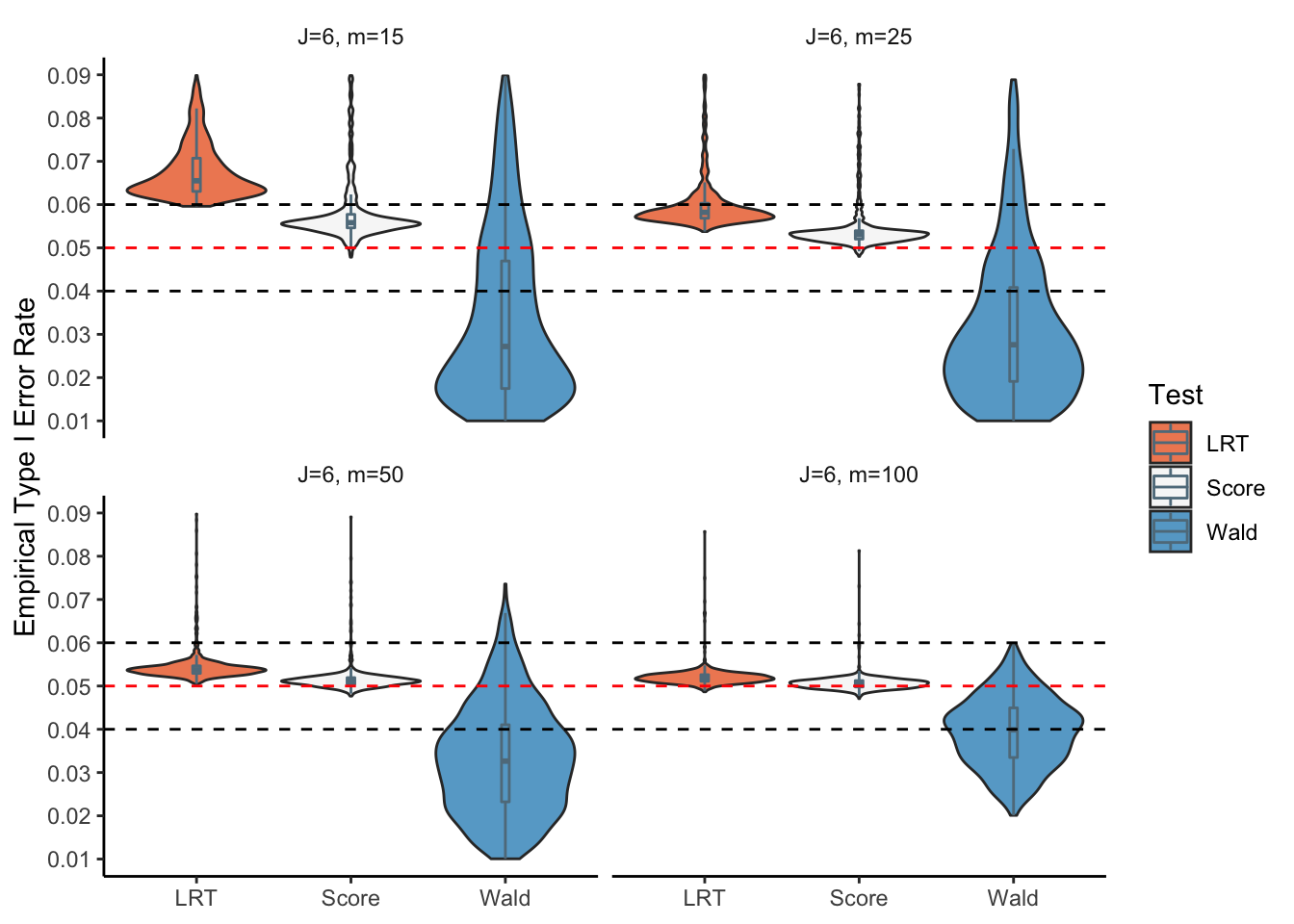}
    \caption{Violin-plots and Boxplots of empirical sizes (J=6)}
    \label{fig:galaxy}
\end{figure}

\begin{figure}[htp]
    \centering
    \includegraphics[width=16cm]{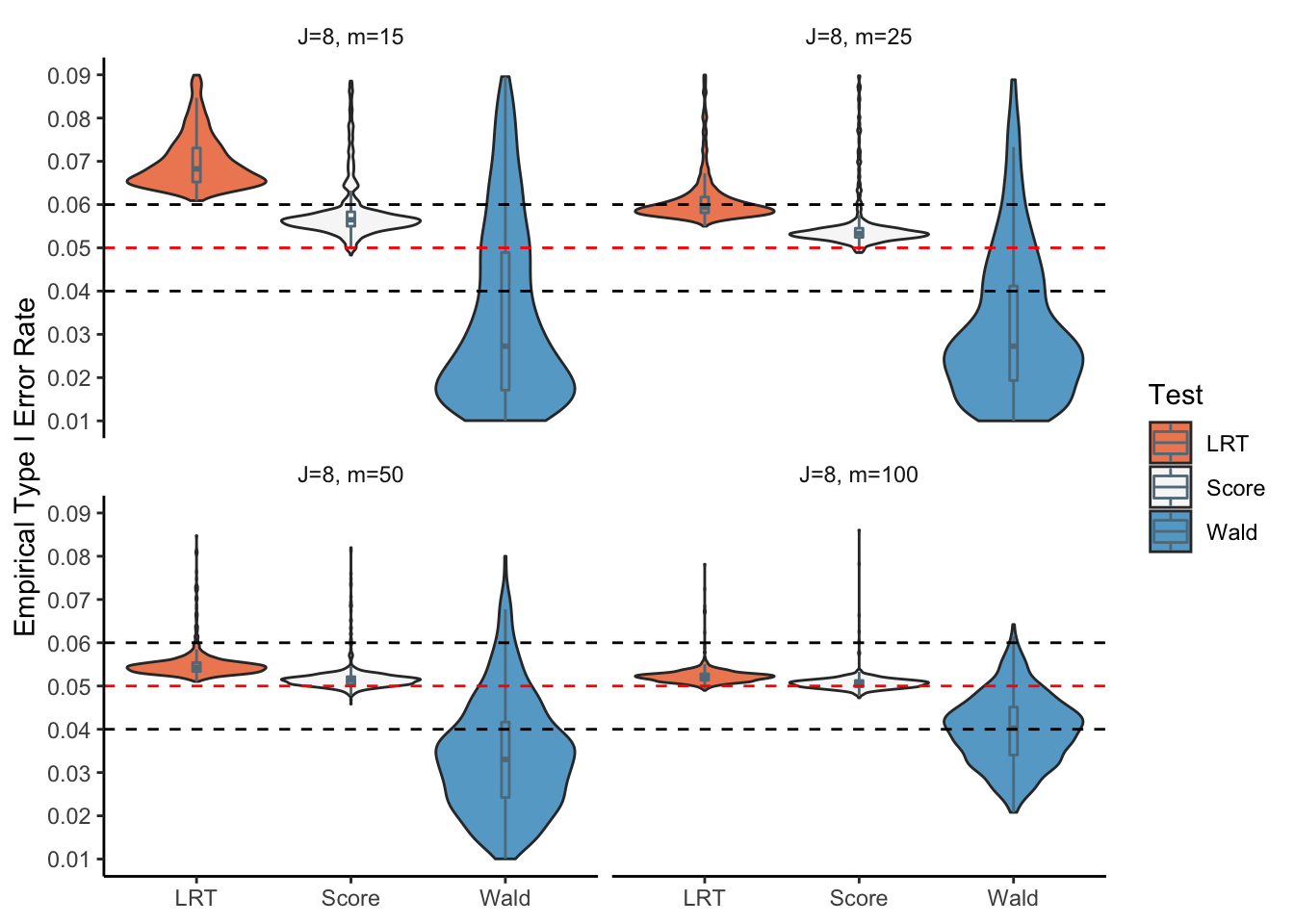}
    \caption{Violin-plots and Boxplots of empirical sizes (J=8)}
    \label{fig:galaxy}
\end{figure}

\subsection{Power}

The third Monte Carlo simulation aims to investigate power performance among three proposed tests. Based on the hypothesis $H_{0}$ : $\delta_{0}$ = 0.5,  and $H_{a}$: $\delta_{a}$ = 1, 1.2, 1.4, we consider the same strata, sample size and parameter setting (Table 1) as the first Monte Carlo simulation of empirical Type I error. As the sample size increases, the powers of the three proposed tests increase for all strata settings (Table 6-9). In addition, as the strata increase, the powers increase for the three proposed tests. Considering empirical Type I error, using the score test is highly recommended.

\begin{singlespace}
\begin{longtable}[t]{rllrrrrrrrrr}
\caption{\label{tab:unnamed-chunk-13}Part of simulation results of the empirical powers for 2 strata. (where $H_{0}$ : $\delta_{0}$ = 0.5, $H_{a}$: $\delta_{a}$ = 1, 1.2, 1.4) }\\
\toprule
\multicolumn{1}{c}{} & \multicolumn{1}{c}{} & \multicolumn{1}{c}{} & \multicolumn{3}{c}{m = 25} & \multicolumn{3}{c}{m = 50} & \multicolumn{3}{c}{m = 100} \\
\cmidrule(l{3pt}r{3pt}){4-6} \cmidrule(l{3pt}r{3pt}){7-9} \cmidrule(l{3pt}r{3pt}){10-12}
$\delta$ & $\gamma$ & $\pi_{1}$ & $T_{L}$ & $T_{SC}$ & $T_{W}$ & $T_{L}$ & $T_{SC}$ & $T_{W}$ & $T_{L}$ & $T_{SC}$ & $T_{W}$\\
\midrule
\endfirsthead
\caption[]{Part of simulation results of the empirical powers for 2 strata.  \textit{(continued)}}\\
\toprule
\multicolumn{1}{c}{} & \multicolumn{1}{c}{} & \multicolumn{1}{c}{} & \multicolumn{3}{c}{m = 25} & \multicolumn{3}{c}{m = 50} & \multicolumn{3}{c}{m = 100} \\
\cmidrule(l{3pt}r{3pt}){4-6} \cmidrule(l{3pt}r{3pt}){7-9} \cmidrule(l{3pt}r{3pt}){10-12}
$\delta$ & $\gamma$ & $\pi_{1}$ & $T_{L}$ & $T_{SC}$ & $T_{W}$ & $T_{L}$ & $T_{SC}$ & $T_{W}$ & $T_{L}$ & $T_{SC}$ & $T_{W}$\\
\midrule
\endhead

\endfoot
\bottomrule
\endlastfoot
1.0 & I & a & 26.63 & 25.19 & 36.64 & 46.61 & 45.74 & 56.84 & 75.30 & 74.88 & 81.43\\
 &  & b & 32.29 & 31.93 & 24.19 & 56.07 & 55.79 & 53.09 & 84.84 & 84.69 & 84.30\\
 &  & c & 24.28 & 23.35 & 24.54 & 41.41 & 40.68 & 45.16 & 68.70 & 68.31 & 72.44\\
 & II & a & 25.87 & 24.38 & 37.96 & 45.10 & 44.27 & 57.17 & 73.99 & 73.62 & 81.19\\
 &  & b & 32.08 & 31.55 & 27.87 & 55.93 & 55.55 & 55.66 & 84.31 & 84.13 & 84.89\\

 &  & c & 23.44 & 22.45 & 26.93 & 40.58 & 39.72 & 46.79 & 68.19 & 67.73 & 73.30\\
 & III & a & 24.79 & 23.33 & 33.68 & 43.01 & 42.11 & 52.49 & 71.06 & 70.55 & 77.60\\
 &  & b & 29.94 & 29.61 & 20.35 & 51.70 & 51.39 & 47.95 & 80.88 & 80.74 & 80.12\\
 &  & c & 22.11 & 21.14 & 20.92 & 38.30 & 37.49 & 41.13 & 65.22 & 64.72 & 68.89\\
 & IV & a & 21.20 & 19.58 & 30.15 & 35.71 & 34.62 & 46.56 & 61.68 & 61.09 & 70.18\\

 &  & b & 25.05 & 24.45 & 16.95 & 43.35 & 42.84 & 40.82 & 71.46 & 71.21 & 71.57\\
 &  & c & 19.23 & 17.91 & 18.62 & 32.05 & 31.04 & 36.04 & 55.53 & 54.87 & 60.50\\
1.2 & I & a & 26.93 & 25.44 & 36.59 & 46.53 & 45.64 & 56.82 & 75.63 & 75.25 & 81.73\\
 &  & b & 32.19 & 31.83 & 24.26 & 55.73 & 55.48 & 52.93 & 84.72 & 84.61 & 84.31\\
 &  & c & 23.58 & 22.64 & 23.71 & 42.03 & 41.27 & 45.80 & 69.50 & 69.02 & 73.13\\

 & II & a & 26.08 & 24.69 & 38.63 & 45.07 & 44.24 & 57.07 & 73.74 & 73.35 & 81.11\\
 &  & b & 32.12 & 31.59 & 28.11 & 55.92 & 55.50 & 55.66 & 84.60 & 84.42 & 85.20\\
 &  & c & 23.21 & 22.10 & 26.80 & 40.38 & 39.53 & 46.57 & 68.18 & 67.67 & 73.16\\
 & III & a & 25.01 & 23.57 & 33.72 & 43.19 & 42.30 & 52.72 & 71.42 & 70.94 & 77.90\\
 &  & b & 29.46 & 29.14 & 20.32 & 51.46 & 51.18 & 47.74 & 80.86 & 80.72 & 80.02\\

 &  & c & 22.52 & 21.55 & 21.26 & 38.34 & 37.56 & 41.36 & 65.27 & 64.77 & 68.83\\
 & IV & a & 21.04 & 19.54 & 30.03 & 35.65 & 34.63 & 46.38 & 61.31 & 60.71 & 69.90\\
 &  & b & 25.19 & 24.60 & 17.21 & 43.64 & 43.05 & 41.07 & 71.28 & 70.95 & 71.32\\
 &  & c & 19.22 & 17.81 & 18.81 & 32.22 & 31.15 & 36.30 & 55.42 & 54.75 & 60.56\\
1.4 & I & a & 41.34 & 40.43 & 54.55 & 68.20 & 68.07 & 78.02 & 92.89 & 92.90 & 95.72\\

 &  & b & 51.00 & 50.73 & 44.36 & 80.03 & 79.85 & 79.33 & 97.66 & 97.64 & 97.77\\
 &  & c & 36.28 & 35.64 & 38.82 & 61.71 & 61.30 & 67.19 & 89.13 & 89.03 & 91.66\\
 & II & a & 40.11 & 39.42 & 55.73 & 66.62 & 66.53 & 78.12 & 91.77 & 91.81 & 95.26\\
 &  & b & 50.43 & 50.02 & 48.41 & 79.41 & 79.20 & 80.61 & 97.64 & 97.61 & 97.94\\
 &  & c & 35.81 & 34.99 & 42.21 & 60.90 & 60.46 & 68.32 & 88.45 & 88.36 & 91.80\\

 & III & a & 38.26 & 37.20 & 50.40 & 64.30 & 64.06 & 74.31 & 90.46 & 90.43 & 93.87\\
 &  & b & 46.88 & 46.61 & 37.68 & 75.52 & 75.34 & 74.04 & 96.33 & 96.29 & 96.36\\
 &  & c & 33.84 & 33.18 & 34.64 & 58.03 & 57.62 & 62.91 & 86.32 & 86.20 & 89.16\\
 & IV & a & 31.58 & 30.29 & 44.33 & 53.92 & 53.44 & 66.05 & 82.66 & 82.62 & 88.62\\
 &  & b & 38.76 & 38.25 & 30.88 & 65.27 & 64.90 & 64.61 & 91.18 & 91.09 & 91.75\\

 &  & c & 28.77 & 27.92 & 30.19 & 48.92 & 48.25 & 55.19 & 78.05 & 77.81 & 82.60\\*
\end{longtable}
\end{singlespace}


\begin{singlespace}
\begin{longtable}[t]{rllrrrrrrrrr}
\caption{\label{tab:unnamed-chunk-13}Part of simulation results of the empirical powers for 4 strata. (where $H_{0}$ : $\delta_{0}$ = 0.5,  $H_{a}$: $\delta_{a}$ = 1, 1.2, 1.4)}\\
\toprule
\multicolumn{1}{c}{} & \multicolumn{1}{c}{} & \multicolumn{1}{c}{} & \multicolumn{3}{c}{m = 25} & \multicolumn{3}{c}{m = 50} & \multicolumn{3}{c}{m = 100} \\
\cmidrule(l{3pt}r{3pt}){4-6} \cmidrule(l{3pt}r{3pt}){7-9} \cmidrule(l{3pt}r{3pt}){10-12}
$\delta$ & $\gamma$ & $\pi_{1}$ & $T_{L}$ & $T_{SC}$ & $T_{W}$ & $T_{L}$ & $T_{SC}$ & $T_{W}$ & $T_{L}$ & $T_{SC}$ & $T_{W}$\\
\midrule
\endfirsthead
\caption[]{Part of simulation results of the empirical powers for 4 strata.  \textit{(continued)}}\\
\toprule
\multicolumn{1}{c}{} & \multicolumn{1}{c}{} & \multicolumn{1}{c}{} & \multicolumn{3}{c}{m = 25} & \multicolumn{3}{c}{m = 50} & \multicolumn{3}{c}{m = 100} \\
\cmidrule(l{3pt}r{3pt}){4-6} \cmidrule(l{3pt}r{3pt}){7-9} \cmidrule(l{3pt}r{3pt}){10-12}
$\delta$ & $\gamma$ & $\pi_{1}$ & $T_{L}$ & $T_{SC}$ & $T_{W}$ & $T_{L}$ & $T_{SC}$ & $T_{W}$ & $T_{L}$ & $T_{SC}$ & $T_{W}$\\
\midrule
\endhead

\endfoot
\bottomrule
\endlastfoot
1.0 & I & a & 32.87 & 30.38 & 48.85 & 59.56 & 58.18 & 72.66 & 89.67 & 89.28 & 94.18\\
 &  & b & 40.86 & 39.69 & 30.64 & 70.84 & 70.34 & 66.69 & 95.60 & 95.51 & 95.31\\
 &  & c & 29.00 & 26.69 & 31.53 & 52.47 & 51.05 & 57.92 & 84.27 & 83.74 & 87.69\\
 & II & a & 32.10 & 29.50 & 51.67 & 57.77 & 56.29 & 73.66 & 88.52 & 88.17 & 94.13\\
 &  & b & 40.33 & 38.83 & 35.05 & 70.27 & 69.52 & 69.56 & 95.53 & 95.38 & 95.77\\

 &  & c & 28.69 & 26.25 & 35.14 & 51.87 & 50.40 & 60.72 & 83.40 & 82.87 & 88.02\\
 & III & a & 30.39 & 27.84 & 44.42 & 55.07 & 53.44 & 68.31 & 86.43 & 85.97 & 91.93\\
 &  & b & 36.86 & 35.63 & 25.39 & 65.70 & 65.09 & 60.02 & 93.41 & 93.23 & 92.82\\
 &  & c & 26.88 & 24.49 & 27.42 & 48.56 & 47.06 & 53.13 & 80.64 & 80.01 & 84.34\\
 & IV & a & 25.34 & 22.30 & 40.20 & 44.79 & 42.91 & 60.59 & 75.83 & 75.03 & 85.05\\

 &  & b & 30.47 & 28.89 & 21.99 & 55.09 & 54.09 & 51.29 & 86.40 & 86.08 & 86.35\\
 &  & c & 22.60 & 19.72 & 25.29 & 39.81 & 38.01 & 46.74 & 70.79 & 69.85 & 76.81\\
1.2 & I & a & 32.75 & 30.27 & 48.53 & 59.52 & 58.08 & 72.99 & 89.65 & 89.28 & 94.25\\
 &  & b & 40.88 & 39.67 & 30.42 & 70.86 & 70.29 & 66.52 & 95.69 & 95.59 & 95.29\\
 &  & c & 28.69 & 26.47 & 31.21 & 51.90 & 50.47 & 57.29 & 84.16 & 83.67 & 87.50\\

 & II & a & 32.00 & 29.35 & 51.70 & 57.66 & 56.26 & 73.73 & 88.27 & 87.86 & 93.85\\
 &  & b & 40.38 & 38.90 & 35.35 & 70.82 & 70.06 & 69.99 & 95.53 & 95.41 & 95.81\\
 &  & c & 28.57 & 26.10 & 35.42 & 51.29 & 49.75 & 59.64 & 83.62 & 83.02 & 88.32\\
 & III & a & 30.26 & 27.61 & 44.43 & 54.79 & 53.27 & 68.30 & 86.34 & 85.84 & 91.86\\
 &  & b & 36.84 & 35.75 & 25.45 & 66.33 & 65.77 & 60.66 & 93.60 & 93.51 & 92.95\\

 &  & c & 26.78 & 24.54 & 27.52 & 48.98 & 47.62 & 53.27 & 80.12 & 79.56 & 83.88\\
 & IV & a & 25.22 & 22.08 & 40.12 & 44.90 & 43.01 & 60.79 & 76.19 & 75.34 & 85.75\\
 &  & b & 30.42 & 28.69 & 22.04 & 55.26 & 54.11 & 51.47 & 86.49 & 86.19 & 86.37\\
 &  & c & 23.01 & 20.06 & 25.44 & 40.13 & 38.27 & 46.62 & 70.89 & 69.92 & 77.00\\
1.4 & I & a & 52.49 & 51.28 & 70.46 & 83.77 & 83.49 & 92.09 & 99.01 & 99.01 & 99.62\\

 &  & b & 65.03 & 63.95 & 56.54 & 92.94 & 92.74 & 92.16 & 99.89 & 99.89 & 99.90\\
 &  & c & 46.87 & 45.12 & 51.77 & 77.37 & 76.73 & 82.95 & 97.77 & 97.72 & 98.60\\
 & II & a & 50.94 & 49.81 & 72.32 & 81.89 & 81.70 & 91.91 & 98.75 & 98.75 & 99.62\\
 &  & b & 64.53 & 63.43 & 61.68 & 92.68 & 92.45 & 93.34 & 99.87 & 99.86 & 99.90\\
 &  & c & 45.21 & 43.42 & 55.03 & 76.33 & 75.69 & 84.21 & 97.49 & 97.43 & 98.67\\

 & III & a & 48.69 & 47.21 & 65.89 & 79.26 & 78.87 & 89.03 & 98.34 & 98.33 & 99.32\\
 &  & b & 59.67 & 58.62 & 48.28 & 89.65 & 89.38 & 87.99 & 99.71 & 99.71 & 99.71\\
 &  & c & 43.01 & 41.14 & 45.97 & 73.60 & 72.82 & 78.78 & 96.65 & 96.55 & 97.90\\
 & IV & a & 39.47 & 37.50 & 58.69 & 68.48 & 67.78 & 82.49 & 94.44 & 94.36 & 97.61\\
 &  & b & 49.51 & 47.94 & 40.46 & 80.65 & 80.14 & 80.02 & 98.54 & 98.50 & 98.76\\

 &  & c & 35.60 & 33.30 & 40.54 & 62.51 & 61.44 & 70.62 & 91.48 & 91.26 & 94.54\\*
\end{longtable}
\end{singlespace}


\begin{singlespace}
\begin{longtable}[t]{rllrrrrrrrrr}
\caption{\label{tab:unnamed-chunk-13}Part of simulation results of the empirical powers for 6 strata. (where $H_{0}$ : $\delta_{0}$ = 0.5, $H_{a}$: $\delta_{a}$ = 1, 1.2, 1.4)}\\
\toprule
\multicolumn{1}{c}{} & \multicolumn{1}{c}{} & \multicolumn{1}{c}{} & \multicolumn{3}{c}{m = 25} & \multicolumn{3}{c}{m = 50} & \multicolumn{3}{c}{m = 100} \\
\cmidrule(l{3pt}r{3pt}){4-6} \cmidrule(l{3pt}r{3pt}){7-9} \cmidrule(l{3pt}r{3pt}){10-12}
$\delta$ & $\gamma$ & $\pi_{1}$ & $T_{L}$ & $T_{SC}$ & $T_{W}$ & $T_{L}$ & $T_{SC}$ & $T_{W}$ & $T_{L}$ & $T_{SC}$ & $T_{W}$\\
\midrule
\endfirsthead
\caption[]{Part of simulation results of the empirical powers for 6 strata.  \textit{(continued)}}\\
\toprule
\multicolumn{1}{c}{} & \multicolumn{1}{c}{} & \multicolumn{1}{c}{} & \multicolumn{3}{c}{m = 25} & \multicolumn{3}{c}{m = 50} & \multicolumn{3}{c}{m = 100} \\
\cmidrule(l{3pt}r{3pt}){4-6} \cmidrule(l{3pt}r{3pt}){7-9} \cmidrule(l{3pt}r{3pt}){10-12}
$\delta$ & $\gamma$ & $\pi_{1}$ & $T_{L}$ & $T_{SC}$ & $T_{W}$ & $T_{L}$ & $T_{SC}$ & $T_{W}$ & $T_{L}$ & $T_{SC}$ & $T_{W}$\\
\midrule
\endhead

\endfoot
\bottomrule
\endlastfoot
1.0 & I & a & 40.02 & 36.70 & 59.66 & 70.92 & 69.37 & 83.71 & 96.27 & 96.04 & 98.35\\
 &  & b & 49.23 & 47.55 & 38.32 & 82.00 & 81.43 & 78.55 & 99.02 & 98.99 & 98.93\\
 &  & c & 34.86 & 31.81 & 39.62 & 62.97 & 61.27 & 69.62 & 92.79 & 92.47 & 95.02\\
 & II & a & 38.51 & 34.82 & 62.47 & 68.57 & 66.99 & 84.25 & 95.63 & 95.39 & 98.35\\
 &  & b & 48.86 & 46.94 & 44.27 & 81.72 & 80.98 & 81.39 & 98.96 & 98.91 & 99.05\\

 &  & c & 34.06 & 30.74 & 44.31 & 61.82 & 59.99 & 71.78 & 92.12 & 91.71 & 95.24\\
 & III & a & 36.85 & 33.41 & 55.33 & 66.00 & 64.15 & 79.57 & 94.25 & 93.94 & 97.20\\
 &  & b & 44.80 & 43.08 & 32.56 & 77.23 & 76.64 & 72.09 & 98.07 & 98.04 & 97.76\\
 &  & c & 32.28 & 29.08 & 35.48 & 58.88 & 57.15 & 64.36 & 90.22 & 89.66 & 92.77\\
 & IV & a & 29.80 & 25.63 & 50.13 & 54.55 & 52.14 & 72.38 & 86.83 & 86.13 & 93.60\\

 &  & b & 36.81 & 34.42 & 28.77 & 66.50 & 65.40 & 62.99 & 94.42 & 94.20 & 94.51\\
 &  & c & 26.98 & 22.90 & 32.89 & 48.50 & 46.17 & 57.24 & 82.32 & 81.45 & 87.58\\
1.2 & I & a & 40.41 & 36.99 & 59.85 & 70.89 & 69.31 & 83.71 & 96.34 & 96.14 & 98.44\\
 &  & b & 49.43 & 47.84 & 38.51 & 81.91 & 81.37 & 78.27 & 98.98 & 98.95 & 98.87\\
 &  & c & 34.68 & 31.62 & 39.72 & 62.86 & 61.16 & 69.42 & 92.86 & 92.50 & 95.11\\

 & II & a & 38.42 & 34.84 & 62.64 & 68.75 & 67.14 & 84.36 & 95.63 & 95.37 & 98.36\\
 &  & b & 48.73 & 46.82 & 44.11 & 82.08 & 81.31 & 81.61 & 98.97 & 98.93 & 99.06\\
 &  & c & 34.28 & 30.91 & 44.00 & 62.16 & 60.39 & 72.05 & 92.31 & 91.88 & 95.44\\
 & III & a & 36.56 & 33.13 & 55.12 & 66.00 & 64.25 & 79.66 & 94.23 & 93.89 & 97.31\\
 &  & b & 45.17 & 43.44 & 32.73 & 77.28 & 76.55 & 72.27 & 98.09 & 98.03 & 97.84\\

 &  & c & 32.09 & 28.92 & 35.37 & 58.70 & 56.87 & 64.17 & 90.36 & 89.91 & 93.00\\
 & IV & a & 30.06 & 25.92 & 50.14 & 54.05 & 51.84 & 72.12 & 86.75 & 86.08 & 93.65\\
 &  & b & 37.00 & 34.67 & 28.79 & 65.97 & 64.87 & 62.56 & 94.60 & 94.41 & 94.51\\
 &  & c & 26.66 & 22.95 & 32.40 & 48.68 & 46.44 & 57.22 & 82.15 & 81.31 & 87.49\\
1.4 & I & a & 63.39 & 61.87 & 81.92 & 92.32 & 92.13 & 97.29 & 99.90 & 99.90 & 99.97\\

 &  & b & 76.50 & 75.35 & 68.99 & 97.88 & 97.78 & 97.57 & 99.99 & 99.99 & 100.00\\
 &  & c & 56.29 & 54.05 & 63.31 & 87.87 & 87.31 & 91.96 & 99.65 & 99.64 & 99.85\\
 & II & a & 61.21 & 59.88 & 82.91 & 91.19 & 91.04 & 97.22 & 99.80 & 99.80 & 99.97\\
 &  & b & 76.17 & 74.95 & 74.00 & 97.78 & 97.67 & 98.08 & 100.00 & 100.00 & 100.00\\
 &  & c & 55.03 & 52.85 & 67.24 & 86.57 & 86.06 & 92.63 & 99.57 & 99.57 & 99.82\\

 & III & a & 58.40 & 56.59 & 77.40 & 89.73 & 89.39 & 96.05 & 99.72 & 99.71 & 99.93\\
 &  & b & 71.22 & 69.95 & 60.57 & 96.20 & 96.05 & 95.46 & 99.98 & 99.98 & 99.98\\
 &  & c & 51.78 & 49.24 & 57.37 & 84.27 & 83.57 & 88.74 & 99.32 & 99.29 & 99.64\\
 & IV & a & 47.65 & 45.07 & 70.12 & 79.89 & 79.25 & 91.34 & 98.62 & 98.59 & 99.58\\
 &  & b & 59.82 & 57.82 & 51.78 & 90.42 & 90.03 & 89.91 & 99.81 & 99.80 & 99.83\\

 &  & c & 42.93 & 39.65 & 51.36 & 74.25 & 73.09 & 82.35 & 97.26 & 97.14 & 98.59\\*
\end{longtable}
\end{singlespace}


\begin{singlespace}
\begin{longtable}[t]{rllrrrrrrrrr}
\caption{\label{tab:unnamed-chunk-13}Part of simulation results of the empirical powers for 8 strata. (where $H_{0}$ : $\delta_{0}$ = 0.5, $H_{a}$: $\delta_{a}$ = 1, 1.2, 1.4)}\\
\toprule
\multicolumn{1}{c}{} & \multicolumn{1}{c}{} & \multicolumn{1}{c}{} & \multicolumn{3}{c}{m = 25} & \multicolumn{3}{c}{m = 50} & \multicolumn{3}{c}{m = 100} \\
\cmidrule(l{3pt}r{3pt}){4-6} \cmidrule(l{3pt}r{3pt}){7-9} \cmidrule(l{3pt}r{3pt}){10-12}
$\delta$ & $\gamma$ & $\pi_{1}$ & $T_{L}$ & $T_{SC}$ & $T_{W}$ & $T_{L}$ & $T_{SC}$ & $T_{W}$ & $T_{L}$ & $T_{SC}$ & $T_{W}$\\
\midrule
\endfirsthead
\caption[]{Part of simulation results of the empirical powers for 8 strata.  \textit{(continued)}}\\
\toprule
\multicolumn{1}{c}{} & \multicolumn{1}{c}{} & \multicolumn{1}{c}{} & \multicolumn{3}{c}{m = 25} & \multicolumn{3}{c}{m = 50} & \multicolumn{3}{c}{m = 100} \\
\cmidrule(l{3pt}r{3pt}){4-6} \cmidrule(l{3pt}r{3pt}){7-9} \cmidrule(l{3pt}r{3pt}){10-12}
$\delta$ & $\gamma$ & $\pi_{1}$ & $T_{L}$ & $T_{SC}$ & $T_{W}$ & $T_{L}$ & $T_{SC}$ & $T_{W}$ & $T_{L}$ & $T_{SC}$ & $T_{W}$\\
\midrule
\endhead

\endfoot
\bottomrule
\endlastfoot
1.0 & I & a & 46.43 & 42.49 & 68.28 & 79.26 & 77.67 & 90.56 & 98.71 & 98.60 & 99.57\\
 &  & b & 57.29 & 55.50 & 45.95 & 89.45 & 89.02 & 86.87 & 99.77 & 99.76 & 99.74\\
 &  & c & 40.76 & 37.04 & 47.42 & 71.99 & 70.35 & 78.53 & 97.00 & 96.80 & 98.21\\
 & II & a & 44.83 & 40.73 & 70.98 & 77.48 & 75.88 & 91.08 & 98.28 & 98.18 & 99.50\\
 &  & b & 56.62 & 54.41 & 52.08 & 89.08 & 88.47 & 88.84 & 99.80 & 99.79 & 99.82\\

 &  & c & 39.56 & 35.53 & 52.08 & 70.52 & 68.76 & 80.60 & 96.51 & 96.24 & 98.19\\
 & III & a & 42.95 & 38.80 & 63.91 & 74.75 & 72.98 & 87.34 & 97.64 & 97.46 & 99.10\\
 &  & b & 52.30 & 50.28 & 39.44 & 85.55 & 84.98 & 81.40 & 99.47 & 99.46 & 99.39\\
 &  & c & 37.17 & 33.43 & 42.08 & 67.15 & 65.29 & 73.34 & 95.29 & 94.93 & 96.91\\
 & IV & a & 34.38 & 29.36 & 58.12 & 62.64 & 60.07 & 80.74 & 93.20 & 92.74 & 97.49\\

 &  & b & 42.94 & 40.16 & 34.89 & 75.04 & 73.84 & 72.37 & 97.78 & 97.70 & 97.80\\
 &  & c & 30.87 & 26.10 & 39.41 & 56.57 & 54.07 & 66.46 & 89.25 & 88.56 & 93.50\\
1.2 & I & a & 46.36 & 42.49 & 68.44 & 79.20 & 77.81 & 90.29 & 98.69 & 98.59 & 99.58\\
 &  & b & 56.96 & 55.08 & 45.89 & 89.38 & 88.99 & 86.72 & 99.78 & 99.77 & 99.73\\
 &  & c & 40.51 & 36.99 & 47.27 & 71.60 & 69.99 & 78.17 & 96.96 & 96.73 & 98.14\\

 & II & a & 44.56 & 40.51 & 70.95 & 77.29 & 75.72 & 90.88 & 98.41 & 98.31 & 99.57\\
 &  & b & 56.35 & 54.11 & 51.92 & 89.20 & 88.63 & 89.02 & 99.76 & 99.75 & 99.78\\
 &  & c & 39.42 & 35.41 & 52.14 & 70.79 & 68.94 & 80.39 & 96.44 & 96.17 & 98.13\\
 & III & a & 42.60 & 38.47 & 63.53 & 74.74 & 73.09 & 87.30 & 97.71 & 97.56 & 99.15\\
 &  & b & 52.36 & 50.39 & 39.68 & 85.43 & 84.78 & 81.17 & 99.46 & 99.44 & 99.37\\

 &  & c & 37.79 & 33.89 & 42.83 & 67.53 & 65.67 & 73.69 & 95.43 & 95.09 & 96.90\\
 & IV & a & 35.08 & 30.14 & 58.73 & 62.74 & 60.31 & 80.71 & 93.00 & 92.52 & 97.36\\
 &  & b & 42.43 & 39.59 & 34.48 & 75.12 & 73.99 & 72.13 & 97.87 & 97.77 & 97.82\\
 &  & c & 30.51 & 25.97 & 39.28 & 56.69 & 54.20 & 66.17 & 89.24 & 88.42 & 93.19\\
1.4 & I & a & 72.00 & 70.38 & 88.71 & 96.50 & 96.38 & 99.08 & 99.99 & 99.99 & 100.00\\

 &  & b & 85.04 & 83.98 & 78.48 & 99.38 & 99.35 & 99.26 & 100.00 & 100.00 & 100.00\\
 &  & c & 64.45 & 62.20 & 72.64 & 93.79 & 93.44 & 96.58 & 99.95 & 99.95 & 99.98\\
 & II & a & 69.89 & 68.35 & 89.87 & 95.82 & 95.69 & 99.16 & 99.98 & 99.98 & 100.00\\
 &  & b & 84.30 & 83.16 & 83.04 & 99.41 & 99.38 & 99.49 & 100.00 & 100.00 & 100.00\\
 &  & c & 63.40 & 60.96 & 76.32 & 92.75 & 92.36 & 96.77 & 99.95 & 99.95 & 99.98\\

 & III & a & 67.11 & 65.17 & 85.14 & 94.76 & 94.57 & 98.32 & 99.97 & 99.97 & 99.99\\
 &  & b & 80.05 & 78.72 & 70.40 & 98.80 & 98.70 & 98.43 & 100.00 & 100.00 & 100.00\\
 &  & c & 60.13 & 57.35 & 66.70 & 91.25 & 90.73 & 94.42 & 99.90 & 99.88 & 99.96\\
 & IV & a & 55.17 & 52.10 & 78.76 & 87.37 & 86.82 & 95.94 & 99.67 & 99.66 & 99.95\\
 &  & b & 68.44 & 66.54 & 61.19 & 95.50 & 95.20 & 95.30 & 99.98 & 99.98 & 99.98\\

 &  & c & 49.91 & 46.09 & 60.05 & 82.62 & 81.56 & 89.40 & 99.13 & 99.09 & 99.64\\*
\end{longtable}

\end{singlespace}

\section{Real Data Example}


We use the double-blinded randomized clinical trial data by Mandel et al. (1982) to illustrate the proposed three tests. In this clinical trial, children who suffer from otitis media with effusion (OME) and simultaneously have bilateral tympanocentesis are randomized into two group: amoxicillin or cefaclor to received the 14-day treatment ~\cite{mandel1982duration}. After the treatment, the number of cured ears are summarized in table 10 with age group as strata. To explore whether the cured rates between two group: amoxicillin and cefaclor among strata age are clinically equivalent, we test the homogeneity based on proposed three tests. The null hypothesis of homogeneity test is $H_{0} = \delta_{1} = \delta_{2} = \delta_{3}$  and alternative hypothesis is
$H_{a}: \delta_{1} \neq \delta_{2} \neq \delta_{3}$, for $J = {1 , 2, 3}$. The MLEs of parameters based on observed data are listed in Table 11, and the three test statistics and $p$-value are summarized in table 12. However, all the $p$-value are greater than 0.05, which means there is no statistical evidence to reject null hypothesis $H_{0} : \delta_{1} = \delta_{2} = \delta_{3}$.

\begin{singlespace}
\begin{longtable}[t]{rrrrrrr}
\caption{\label{tab:unnamed-chunk-2} Frequency of number of OME-free ears after treatment.}\\
\toprule
\multicolumn{1}{c}{} & \multicolumn{6}{c}{Age group} \\
\cmidrule(l{3pt}r{3pt}){2-7}
\multicolumn{1}{c}{} & \multicolumn{2}{c}{ $<$2 years} & \multicolumn{2}{c}{2-5 years} & \multicolumn{2}{c}{$\geq$6 years} \\
\cmidrule(l{3pt}r{3pt}){2-3} \cmidrule(l{3pt}r{3pt}){4-5} \cmidrule(l{3pt}r{3pt}){6-7}
Number of OME-Free ears & Cefactor & Amoxicillin & Cefactor & Amoxicillin & Cefactor & Amoxicillin\\
\midrule
\endfirsthead
\caption[]{Table 1. Frequency of number of OME-free ears after treatment. \textit{(continued)}}\\
\toprule
\multicolumn{1}{c}{} & \multicolumn{6}{c}{Age group} \\
\cmidrule(l{3pt}r{3pt}){2-7}
\multicolumn{1}{c}{} & \multicolumn{2}{c}{<2 years} & \multicolumn{2}{c}{2-5 years} & \multicolumn{2}{c}{$\geq$6 years} \\
\cmidrule(l{3pt}r{3pt}){2-3} \cmidrule(l{3pt}r{3pt}){4-5} \cmidrule(l{3pt}r{3pt}){6-7}
Number of OME-Free ears & Cefactor & Amoxicillin & Cefactor & Amoxicillin & Cefactor & Amoxicillin\\
\midrule
\endhead

\endfoot
\bottomrule
\endlastfoot
0 & 8 & 11 & 6 & 3 & 0 & 1\\
1 & 2 & 2 & 6 & 1 & 1 & 0\\
2 & 8 & 2 & 10 & 5 & 3 & 6\\*
\end{longtable}
\end{singlespace}

\begin{singlespace}
\begin{longtable}[t]{lrrrrrl}
\caption{\label{tab:unnamed-chunk-4}MlEs of parameters based on observed data.}\\
\toprule
\multicolumn{1}{c}{} & \multicolumn{3}{c}{Global MlEs} & \multicolumn{3}{c}{Unconstrained MLEs} \\
\cmidrule(l{3pt}r{3pt}){2-4} \cmidrule(l{3pt}r{3pt}){5-7}
\multicolumn{1}{l}{Age groups} & \multicolumn{1}{l}{$\tilde{\pi}_{1}$} & \multicolumn{1}{l}{$\tilde{\gamma}$} & \multicolumn{1}{l}{$\tilde{\delta}$} & \multicolumn{1}{l}{$\hat{\pi}_{1}$} & \multicolumn{1}{l}{$\hat{\gamma}$} & \multicolumn{1}{l}{$\hat{\delta}$}\\
\midrule
\endfirsthead
\caption[]{MlEs of parameters based on observed data. \textit{(continued)}}\\
\toprule
\multicolumn{1}{c}{} & \multicolumn{3}{c}{Global MlEs} & \multicolumn{3}{c}{Unconstrained MLEs} \\
\cmidrule(l{3pt}r{3pt}){2-4} \cmidrule(l{3pt}r{3pt}){5-7}
\multicolumn{1}{l}{Age groups} & \multicolumn{1}{l}{$\tilde{\pi}_{1}$} & \multicolumn{1}{l}{$\tilde{\gamma}$} & \multicolumn{1}{l}{$\tilde{\delta}$} & \multicolumn{1}{l}{$\hat{\pi}_{1}$} & \multicolumn{1}{l}{$\hat{\gamma}$} & \multicolumn{1}{l}{$\hat{\delta}$}\\
\midrule
\endhead

\endfoot
\bottomrule
\endlastfoot
\multicolumn{1}{l}{$<$2 years} & \multicolumn{1}{l}{0.4762} & \multicolumn{1}{l}{0.8333} & \multicolumn{1}{l}{0.4800} & \multicolumn{1}{l}{0.4036} & \multicolumn{1}{l}{0.8333} & \multicolumn{1}{l}{0.8174}\\
\multicolumn{1}{l}{2-5 years} & \multicolumn{1}{l}{0.6116} & \multicolumn{1}{l}{0.8108} & \multicolumn{1}{l}{0.9167} & \multicolumn{1}{l}{0.6249} & \multicolumn{1}{l}{0.8108} & \multicolumn{1}{l}{-}\\
\multicolumn{1}{l}{$\geq$6 years} & \multicolumn{1}{l}{0.9500} & \multicolumn{1}{l}{0.9474} & \multicolumn{1}{l}{0.8572} & \multicolumn{1}{l}{0.9500} & \multicolumn{1}{l}{0.9474} & \multicolumn{1}{l}{-}\\*
\end{longtable}

\end{singlespace}

\begin{singlespace}
\begin{longtable}[t]{lrrr}
\caption{\label{tab:unnamed-chunk-5}The values of statistics and p-values for three different test.}\\
\toprule
 & $T_{L}$ & $T_{SC}$ & $T_{W}$\\
\midrule
\endfirsthead
\caption[]{The values of statistics and p-values for three different test. \textit{(continued)}}\\
\toprule
 & $T_{L}$ & $T_{SC}$ & $T_{W}$\\
\midrule
\endhead

\endfoot
\bottomrule
\endlastfoot
Statistic & 1.6918 & 1.6392 & 2.3520\\
P & 0.4292 & 0.4406 & 0.3085\\*
\end{longtable}
\end{singlespace}

\section{Conclusions}

This article utilizes three MLE-based tests (LRT, Wald-type test, and score test) to test the homogeneity relative risk of two proportions on stratified bilateral correlated data.
  
 Three Monte Carlo simulation results show that the score test yields a robust performance by empirical Type I error and power. Even with a small sample size and multiple strata, the score test still generates stable empirical Type I error and satisfactory power. Meanwhile, the likelihood ratio test and Wald-type test offer reasonable power, but comes with unstable empirical Type I error performances. By incorporating a small sample size, the Wald-type test shows unsatisfactory performance of empirical Type I error, but the performance tends to be reasonable by increasing the sample size. Under a small sample size with multiple strata, the score test is slightly unstable, but the performance also tends to be robust by increasing the sample size.
 
 However, asymptotic methods might have some limitations because of poor performance under a small sample size with multiple strata. Future work might consider exact tests to investigate related issues.

\section{Appendix}

\subsection{Derivation of score statistic}
The first order differential equations of stratum j are:

$$\frac{\partial l_{j}}{\partial \delta}=
\frac{ m_{12j} }{ \delta}+\frac{ m_{22j} }{ \delta}+\frac{ m_{02j} \, \pi_{1 j } \,{\left(\gamma_j -2\right)}}{ \delta\, \pi_{1 j } \,{\left(\gamma_j -2\right)}+1} ,$$

$$\frac{\partial l_{j}}{\partial \pi_{1 j}} = \frac{ m_{12j} }{ \pi_{1 j } }+\frac{ m_{21j} }{ \pi_{1 j } }+\frac{ m_{22j} }{ \pi_{1 j } }+\frac{ m_{01j} \,{\left(\gamma_j -2\right)}}{ \pi_{1 j } \,{\left(\gamma_j -2\right)}+1}+\frac{ \delta\, m_{02j} \,{\left(\gamma_j -2\right)}}{ \delta\, \pi_{1 j } \,{\left(\gamma_j -2\right)}+1}+\frac{ m_{11j} \,{\left(2\,\gamma_j -2\right)}}{2\, \pi_{1 j } \,{\left(\gamma_j -1\right)}} , $$

$$\frac{\partial l_{j}}{\partial \gamma_{j}} = \frac{ m_{21j} }{\gamma_j }+\frac{ m_{22j} }{\gamma_j }+\frac{ m_{11j} }{\gamma_j -1}+\frac{ m_{12j} }{\gamma_j -1}+\frac{ m_{01j} \, \pi_{1 j } }{ \pi_{1 j } \,{\left(\gamma_j -2\right)}+1}+\frac{ \delta\, m_{02j} \, \pi_{1 j } }{ \delta\, \pi_{1 j } \,{\left(\gamma_j -2\right)}+1} . $$

The second order differential equations of stratum j are:

$$ \frac{\partial^{2} l_{j}}{\partial \delta^{2}} = 
-\frac{ m_{12j} }{ \delta^2 }-\frac{ m_{22j} }{ \delta^2 }-\frac{ m_{02j} \,{ \pi_{1 j } }^2 \,{{\left(\gamma_j -2\right)}}^2 }{{{\left( \delta \, \pi_{1 j } \,{\left(\gamma_j -2\right)}+1\right)}}^2 } , $$

$$\frac{\partial^{2} l_{j}}{\partial \delta \partial \pi_{1 j}}=
\frac{ m_{02j} \,{\left(\gamma_j -2\right)}}{ \delta\, \pi_{1 j } \,{\left(\gamma_j -2\right)}+1}-\frac{ \delta\, m_{02j} \, \pi_{1 j } \,{{\left(\gamma_j -2\right)}}^2 }{{{\left( \delta\, \pi_{1 j } \,{\left(\gamma_j -2\right)}+1\right)}}^2 } , $$

$$\frac{\partial^{2} l_{j}}{\partial \delta \partial \gamma_{j}}= \frac{ m_{02j} \, \pi_{1 j } }{ \delta\, \pi_{1 j } \,{\left(\gamma_j -2\right)}+1}-\frac{ \delta\, m_{02j} \,{ \pi_{1 j } }^2 \,{\left(\gamma_j -2\right)}}{{{\left( \delta\, \pi_{1 j } \,{\left(\gamma_j -2\right)}+1\right)}}^2 } , $$

$$\frac{\partial^{2} l_{j}}{ \partial \pi_{1 j} \partial \delta} =\frac{ m_{02j} \,{\left(\gamma_j -2\right)}}{ \delta\, \pi_{1 j } \,{\left(\gamma_j -2\right)}+1}-\frac{ \delta\, m_{02j} \, \pi_{1 j } \,{{\left(\gamma_j -2\right)}}^2 }{{{\left( \delta\, \pi_{1 j } \,{\left(\gamma_j -2\right)}+1\right)}}^2 } , $$

$$\frac{\partial^{2} l_{j}}{\partial \pi_{1 j}^{2}} = -\frac{ m_{12j} }{{ \pi_{1 j } }^2 }-\frac{ m_{21j} }{{ \pi_{1 j } }^2 }-\frac{ m_{22j} }{{ \pi_{1 j } }^2 }-\frac{ m_{01j} \,{{\left(\gamma_j -2\right)}}^2 }{{{\left( \pi_{1 j } \,{\left(\gamma_j -2\right)}+1\right)}}^2 }-\frac{\delta^2 \, m_{02j} \,{{\left(\gamma_j -2\right)}}^2 }{{{\left( \delta\, \pi_{1 j } \,{\left(\gamma_j -2\right)}+1\right)}}^2 }-\frac{ m_{11j} \,{\left(2\,\gamma_j -2\right)}}{2\,{ \pi_{1 j } }^2 \,{\left(\gamma_j -1\right)}} , $$

$$\frac{\partial^{2} l_{j}}{\partial \pi_{1 j} \partial \gamma_{j}} = \frac{ m_{01j} }{{{\left( \pi_{1 j } \,{\left(\gamma_j -2\right)}+1\right)}}^2 }+\frac{ \delta\, m_{02j} }{{{\left( \delta\, \pi_{1 j } \,{\left(\gamma_j -2\right)}+1\right)}}^2 } , $$

$$\frac{\partial^{2} l_{j}}{\partial \gamma_{j} \partial \delta } =\frac{ m_{02j} \, \pi_{1 j } }{ \delta\, \pi_{1 j } \,{\left(\gamma_j -2\right)}+1}-\frac{ \delta\, m_{02j} \,{ \pi_{1 j } }^2 \,{\left(\gamma_j -2\right)}}{{{\left( \delta\, \pi_{1 j } \,{\left(\gamma_j -2\right)}+1\right)}}^2 }, $$

$$\frac{\partial^{2} l_{j}}{\partial \gamma_{j} \partial \pi_{1 j} }
\frac{ m_{01j} }{{{\left( \pi_{1 j } \,{\left(\gamma_j -2\right)}+1\right)}}^2 }+\frac{ \delta\, m_{02j} }{{{\left( \pi_{1 j } \,{\left(2\, \delta- \delta\,\gamma_j \right)}-1\right)}}^2 } , $$

$$\frac{\partial^{2} l_{j}}{\partial \gamma_{ j}^{2}} =-\frac{ m_{21j} }{{\gamma_j }^2 }-\frac{ m_{22j} }{{\gamma_j }^2 }-\frac{ m_{11j} }{{{\left(\gamma_j -1\right)}}^2 }-\frac{ m_{12j} }{{{\left(\gamma_j -1\right)}}^2 }-\frac{ m_{01j} \,{ \pi_{1 j } }^2 }{{{\left( \pi_{1 j } \,{\left(\gamma_j -2\right)}+1\right)}}^2 }-\frac{\delta^2 \, m_{02j} \,{ \pi_{1 j } }^2 }{{{\left( \delta\, \pi_{1 j } \,{\left(\gamma_j -2\right)}+1\right)}}^2 } . $$








Fisher information of stratum j are: 

$$I_{11}^{(j)}=E\left(-\frac{\partial^{2} l_{j}}{\partial \delta^{2}}\right) = \frac{m_{+2j} }{ \delta^2 \,{\left( \delta \, \pi_{1 j } \,{\left(\gamma_j -2\right)}+1\right)}}-\frac{m_{+2j} }{ \delta^2 } , $$

$$I_{12}^{(j)}=I_{21}^{(j)}=E\left(-\frac{\partial^{2} l_{j}}{\partial \delta \partial \pi_{1 j}}\right)= \-\frac{m_{+2j} \,{\left(\gamma_j -2\right)}}{ \delta \, \pi_{1 j } \,{\left(\gamma_j -2\right)}+1} , $$

$$I_{13}^{(j)}=I_{31}^{(j)}=E\left(-\frac{\partial^{2} l_{j}}{\partial \delta \partial  \gamma_{j}}\right) = -\frac{m_{+2j} \, \pi_{1 j } }{ \delta \, \pi_{1 j } \,{\left(\gamma_j -2\right)}+1} , $$

$$I_{22}^{(j)}=E\left(-\frac{\partial^{2} l_{j}}{\partial \pi_{1 j}^{2}}\right)= \frac{m_{+1j} \,{{\left(\gamma_j -2\right)}}^2 }{ \pi_{1 j } \,{\left(\gamma_j -2\right)}+1}-\frac{{\left(\gamma_j -2\right)}\,{\left(m_{+1j} + \delta \,m_{+2j} \right)}}{ \pi_{1 j } }-\frac{ \delta^2 \,m_{+2j} \,{{\left(\gamma_j -2\right)}}^2 }{ \pi_{1 j } \,{\left(2\, \delta - \delta \,\gamma_j \right)}-1} , $$

$$I_{23}^{(j)}=I_{32}^{(j)}=E\left(-\frac{\partial^{2} l_{j}}{\partial \pi_{1 j} \gamma_{j}}\right) = \frac{ \delta \,m_{+2j} }{ \pi_{1 j } \,{\left(2\, \delta - \delta \,\gamma_j \right)}-1}-\frac{m_{+1j} }{ \pi_{1 j } \,{\left(\gamma_j -2\right)}+1} , $$

$$
\begin{aligned}
I_{33}^{(j)}=E\left(-\frac{\partial^{2} l_{j}}{\partial \gamma_{j}^{2}}\right) 
& = \frac{m_{+1j} \, \pi_{1 j } }{\gamma_j }
-\frac{2\,m_{+1j} \, \pi_{1 j } }{\gamma_j -1}
+\frac{m_{+1j} \,{ \pi_{1 j } }^2 }{ \pi_{1 j } \,{\left(\gamma_j -2\right)}+1} \\
& +\frac{ \delta^2 \,m_{+2j} \,{ \pi_{1 j } }^2 }{ \delta \, \pi_{1 j } \,{\left(\gamma_j -2\right)}+1}+\frac{ \delta \,m_{+2j} \, \pi_{1 j } }{\gamma_j }-\frac{2\, \delta \,m_{+2j} \, \pi_{1 j } }{\gamma_j -1} . 
\end{aligned}
$$

\subsection{Derivation of Wald-type statistic}

The first order differential equations of stratum j are:

$$\frac{\partial l_{j}}{\partial \delta_{j}}=
\frac{ m_{12j} }{ \delta_{j}}+\frac{ m_{22j} }{ \delta_{j}}+\frac{ m_{02j} \, \pi_{1 j } \,{\left(\gamma_j -2\right)}}{ \delta_{j}\, \pi_{1 j } \,{\left(\gamma_j -2\right)}+1} , $$

$$\frac{\partial l_{j}}{\partial \pi_{1 j}} = \frac{ m_{12j} }{ \pi_{1 j } }+\frac{ m_{21j} }{ \pi_{1 j } }+\frac{ m_{22j} }{ \pi_{1 j } }+\frac{ m_{01j} \,{\left(\gamma_j -2\right)}}{ \pi_{1 j } \,{\left(\gamma_j -2\right)}+1}+\frac{ \delta_{j}\, m_{02j} \,{\left(\gamma_j -2\right)}}{ \delta_{j}\, \pi_{1 j } \,{\left(\gamma_j -2\right)}+1}+\frac{ m_{11j} \,{\left(2\,\gamma_j -2\right)}}{2\, \pi_{1 j } \,{\left(\gamma_j -1\right)}} , $$

$$\frac{\partial l_{j}}{\partial \gamma_{j}} = \frac{ m_{21j} }{\gamma_j }+\frac{ m_{22j} }{\gamma_j }+\frac{ m_{11j} }{\gamma_j -1}+\frac{ m_{12j} }{\gamma_j -1}+\frac{ m_{01j} \, \pi_{1 j } }{ \pi_{1 j } \,{\left(\gamma_j -2\right)}+1}+\frac{ \delta_{j}\, m_{02j} \, \pi_{1 j } }{ \delta_{j}\, \pi_{1 j } \,{\left(\gamma_j -2\right)}+1} . $$

The second order differential equations of stratum j are:

$$ \frac{\partial^{2} l_{j}}{\partial \delta_{j}^{2}} = 
-\frac{ m_{12j} }{ \delta_{j}^2 }-\frac{ m_{22j} }{ \delta_{j}^2 }-\frac{ m_{02j} \,{ \pi_{1 j } }^2 \,{{\left(\gamma_j -2\right)}}^2 }{{{\left( \delta_{j} \, \pi_{1 j } \,{\left(\gamma_j -2\right)}+1\right)}}^2 } , $$

$$\frac{\partial^{2} l_{j}}{\partial \delta_{j} \partial \pi_{1 j}}=
\frac{ m_{02j} \,{\left(\gamma_j -2\right)}}{ \delta_{j}\, \pi_{1 j } \,{\left(\gamma_j -2\right)}+1}-\frac{ \delta_{j}\, m_{02j} \, \pi_{1 j } \,{{\left(\gamma_j -2\right)}}^2 }{{{\left( \delta_{j}\, \pi_{1 j } \,{\left(\gamma_j -2\right)}+1\right)}}^2 } , $$

$$\frac{\partial^{2} l_{j}}{\partial \delta_{j} \partial \gamma_{j}}= \frac{ m_{02j} \, \pi_{1 j } }{ \delta_{j}\, \pi_{1 j } \,{\left(\gamma_j -2\right)}+1}-\frac{ \delta_{j}\, m_{02j} \,{ \pi_{1 j } }^2 \,{\left(\gamma_j -2\right)}}{{{\left( \delta_{j}\, \pi_{1 j } \,{\left(\gamma_j -2\right)}+1\right)}}^2 } , $$

$$\frac{\partial^{2} l_{j}}{ \partial \pi_{1 j} \partial \delta_{j}} =\frac{ m_{02j} \,{\left(\gamma_j -2\right)}}{ \delta_{j}\, \pi_{1 j } \,{\left(\gamma_j -2\right)}+1}-\frac{ \delta_{j}\, m_{02j} \, \pi_{1 j } \,{{\left(\gamma_j -2\right)}}^2 }{{{\left( \delta_{j}\, \pi_{1 j } \,{\left(\gamma_j -2\right)}+1\right)}}^2 } , $$

$$\frac{\partial^{2} l_{j}}{\partial \pi_{1 j}^{2}} = -\frac{ m_{12j} }{{ \pi_{1 j } }^2 }-\frac{ m_{21j} }{{ \pi_{1 j } }^2 }-\frac{ m_{22j} }{{ \pi_{1 j } }^2 }-\frac{ m_{01j} \,{{\left(\gamma_j -2\right)}}^2 }{{{\left( \pi_{1 j } \,{\left(\gamma_j -2\right)}+1\right)}}^2 }-\frac{\delta^2 \, m_{02j} \,{{\left(\gamma_j -2\right)}}^2 }{{{\left( \delta_{j}\, \pi_{1 j } \,{\left(\gamma_j -2\right)}+1\right)}}^2 }-\frac{ m_{11j} \,{\left(2\,\gamma_j -2\right)}}{2\,{ \pi_{1 j } }^2 \,{\left(\gamma_j -1\right)}} , $$

$$\frac{\partial^{2} l_{j}}{\partial \pi_{1 j} \partial \gamma_{j}} = \frac{ m_{01j} }{{{\left( \pi_{1 j } \,{\left(\gamma_j -2\right)}+1\right)}}^2 }+\frac{ \delta_{j}\, m_{02j} }{{{\left( \delta_{j}\, \pi_{1 j } \,{\left(\gamma_j -2\right)}+1\right)}}^2 } , $$

$$\frac{\partial^{2} l_{j}}{\partial \gamma_{j} \partial \delta_{j} } =\frac{ m_{02j} \, \pi_{1 j } }{ \delta_{j}\, \pi_{1 j } \,{\left(\gamma_j -2\right)}+1}-\frac{ \delta_{j}\, m_{02j} \,{ \pi_{1 j } }^2 \,{\left(\gamma_j -2\right)}}{{{\left( \delta_{j}\, \pi_{1 j } \,{\left(\gamma_j -2\right)}+1\right)}}^2 } , $$

$$\frac{\partial^{2} l_{j}}{\partial \gamma_{j} \partial \pi_{1 j} }
\frac{ m_{01j} }{{{\left( \pi_{1 j } \,{\left(\gamma_j -2\right)}+1\right)}}^2 }+\frac{ \delta_{j}\, m_{02j} }{{{\left( \pi_{1 j } \,{\left(2\, \delta_{j}- \delta_{j}\,\gamma_j \right)}-1\right)}}^2 } , $$

$$\frac{\partial^{2} l_{j}}{\partial \gamma_{ j}^{2}} =-\frac{ m_{21j} }{{\gamma_j }^2 }-\frac{ m_{22j} }{{\gamma_j }^2 }-\frac{ m_{11j} }{{{\left(\gamma_j -1\right)}}^2 }-\frac{ m_{12j} }{{{\left(\gamma_j -1\right)}}^2 }-\frac{ m_{01j} \,{ \pi_{1 j } }^2 }{{{\left( \pi_{1 j } \,{\left(\gamma_j -2\right)}+1\right)}}^2 }-\frac{\delta^2 \, m_{02j} \,{ \pi_{1 j } }^2 }{{{\left( \delta_{j}\, \pi_{1 j } \,{\left(\gamma_j -2\right)}+1\right)}}^2 } . $$








Fisher information of stratum j are: 

$$I_{W11}^{(j)}=E\left(-\frac{\partial^{2} l_{j}}{\partial \delta_{j}^{2}}\right) = \frac{m_{+2j} }{ \delta_{j}^2 \,{\left( \delta_{j} \, \pi_{1 j } \,{\left(\gamma_j -2\right)}+1\right)}}-\frac{m_{+2j} }{ \delta_{j}^2 } , $$

$$I_{W12}^{(j)}=I_{W21}^{(j)}=E\left(-\frac{\partial^{2} l_{j}}{\partial \delta_{j} \partial \pi_{1 j}}\right)= \-\frac{m_{+2j} \,{\left(\gamma_j -2\right)}}{ \delta_{j} \, \pi_{1 j } \,{\left(\gamma_j -2\right)}+1} , $$

$$I_{W13}^{(j)}=I_{W31}^{(j)}=E\left(-\frac{\partial^{2} l_{j}}{\partial \delta_{j} \partial  \gamma_{j}}\right) = -\frac{m_{+2j} \, \pi_{1 j } }{ \delta_{j} \, \pi_{1 j } \,{\left(\gamma_j -2\right)}+1} , $$

$$I_{W22}^{(j)}=E\left(-\frac{\partial^{2} l_{j}}{\partial \pi_{1 j}^{2}}\right)= \frac{m_{+1j} \,{{\left(\gamma_j -2\right)}}^2 }{ \pi_{1 j } \,{\left(\gamma_j -2\right)}+1}-\frac{{\left(\gamma_j -2\right)}\,{\left(m_{+1j} + \delta_{j} \,m_{+2j} \right)}}{ \pi_{1 j } }-\frac{ \delta_{j}^2 \,m_{+2j} \,{{\left(\gamma_j -2\right)}}^2 }{ \pi_{1 j } \,{\left(2\, \delta_{j} - \delta_{j} \,\gamma_j \right)}-1} , $$

$$I_{W23}^{(j)}=I_{W32}^{(j)}=E\left(-\frac{\partial^{2} l_{j}}{\partial \pi_{1 j} \gamma_{j}}\right) = \frac{ \delta_{j} \,m_{+2j} }{ \pi_{1 j } \,{\left(2\, \delta_{j} - \delta_{j} \,\gamma_j \right)}-1}-\frac{m_{+1j} }{ \pi_{1 j } \,{\left(\gamma_j -2\right)}+1} , $$

$$
\begin{aligned}
I_{W33}^{(j)}=E\left(-\frac{\partial^{2} l_{j}}{\partial \gamma_{j}^{2}}\right) 
& = \frac{m_{+1j} \, \pi_{1 j } }{\gamma_j }
-\frac{2\,m_{+1j} \, \pi_{1 j } }{\gamma_j -1}
+\frac{m_{+1j} \,{ \pi_{1 j } }^2 }{ \pi_{1 j } \,{\left(\gamma_j -2\right)}+1} \\
& +\frac{ \delta_{j}^2 \,m_{+2j} \,{ \pi_{1 j } }^2 }{ \delta_{j} \, \pi_{1 j } \,{\left(\gamma_j -2\right)}+1}
+\frac{ \delta_{j} \,m_{+2j} \, \pi_{1 j } }{\gamma_j }
-\frac{2\, \delta_{j} \,m_{+2j} \, \pi_{1 j } }{\gamma_j -1}. 
\end{aligned}
$$

\bibliographystyle{unsrt}
\bibliography{1.bib}
\end{document}